\documentclass[12pt]{article}

\usepackage{fixltx2e}
\usepackage{textcomp}
\usepackage{fullpage}
\usepackage{amsfonts}
\usepackage{verbatim}
\usepackage[english]{babel}
\usepackage{pifont}
\usepackage{color}
\usepackage{setspace}
\usepackage{lscape}
\usepackage{indentfirst}
\usepackage[normalem]{ulem}
\usepackage{booktabs}
\usepackage{natbib}
\usepackage{float}
\usepackage{latexsym}
\usepackage{hyperref}
\usepackage{url}
\usepackage{epsfig}
\usepackage{graphicx}
\usepackage{amssymb}
\usepackage{amsmath}
\usepackage{bm}
\usepackage{array}
\usepackage{ifthen}
\usepackage{caption}
\usepackage{xcolor}
\usepackage{amsthm}
\usepackage{amstext}
\usepackage{nicefrac}
\usepackage{algorithm}
\usepackage{algorithmic}
\usepackage[scientific-notation=true]{siunitx}
\usepackage{subfigure}
\usepackage[flushleft]{threeparttable}
\usepackage{lineno}
\usepackage{adjustbox}

\newcommand{\nL}{n_l} 
\newcommand{\nS}{n} 
\newcommand{\nI}{n_i} 

\newcommand{\BayesianMultispeciesCoalescent}{*BEAST}
\newcommand{\BayesianSupermatrix}{Bayesian supermatrix}
\newcommand{\MLSupermatrix}{Maximum-likelihood supermatrix}
\newcommand{\MPEST}{\mbox{MP-EST}}

\raggedright
\renewcommand{\section}[1]{%
\bigskip
\begin{center}
\begin{Large}
\normalfont\scshape #1
\medskip
\end{Large}
\end{center}}

\renewcommand{\subsection}[1]{%
\bigskip
\begin{center}
\begin{large}
\normalfont\itshape #1
\end{large}
\end{center}}

\renewcommand{\subsubsection}[1]{%
\vspace{2ex}
\noindent
\textit{#1.}---}

\renewcommand{\tableofcontents}{}

\bibpunct{(}{)}{;}{a}{}{,}

\begin{document}
\begin{flushright}
Version dated: \today
\end{flushright}
\bigskip

\noindent Performance \& Accuracy of Species Tree Methods

\bigskip
\medskip
\begin{center}

\noindent{\Large \bf Computational Performance and Statistical Accuracy of *BEAST and Comparisons with Other Methods}
\bigskip


\noindent {\normalsize \sc Huw A. Ogilvie$^{1}$, Joseph Heled$^{2,3}$, Dong Xie$^{2,3}$ and Alexei J. Drummond$^{2,3}$}
\noindent {\small \it $^1$Evolution, Ecology and Genetics, Research School of Biology, The Australian National University, Canberra, Australia;\\
$^2$Department of Computer Science, University of Auckland, Auckland, New Zealand;\\
$^3$Allan Wilson Centre for Molecular Ecology and Evolution, University of Auckland, Auckland, New Zealand}
\end{center}
\medskip
\noindent{\bf Corresponding author:} Alexei J. Drummond,
Department of Computer Science, University of Auckland, Auckland, New Zealand;
E-mail: alexei@cs.auckland.ac.nz\\

\vspace{3in}

\subsubsection{Abstract} Under the multispecies coalescent model of molecular
evolution, gene trees have independent evolutionary histories within a shared species tree. 
In comparison, supermatrix
concatenation methods assume that gene trees share a single common genealogical history, thereby equating
gene coalescence with species divergence. The multispecies coalescent is
supported by previous studies which found that its predicted distributions fit
empirical data, and that concatenation is not a consistent estimator of the
species tree. *BEAST, a fully Bayesian implementation of the multispecies
coalescent, is popular but computationally intensive, so the increasing size of
phylogenetic data sets is both a computational challenge and an opportunity for
better systematics. Using simulation studies, we characterize the scaling
behaviour of *BEAST, and enable quantitative prediction of the impact increasing
the number of loci has on both computational performance and statistical
accuracy. Follow up simulations over a wide range of parameters show that the
statistical performance of *BEAST relative to concatenation improves both as
branch length is reduced and as the number of loci is increased. Finally, using
simulations based on estimated parameters from two phylogenomic data sets, we
compare the performance of a range of species tree and concatenation methods to
show that using *BEAST with tens of loci can be preferable to using
concatenation with thousands of loci. Our results provide insight into the
practicalities of Bayesian species tree estimation, the number of loci required
to obtain a given level of accuracy and the situations in which supermatrix or
summary methods will be outperformed by the fully Bayesian multispecies
coalescent.

\noindent(Keywords: Phylogenomics, Species tree, Gene tree, Bayesian
phylogenetics, Multispecies coalescent, Concatenation, Supermatrix)

\vspace{1.5in}

\section{Introduction}

In recent years a number of new techniques have applied next-generation
sequencing to phylogenetics and phylogeography \citep{McCormack:2013kx}. These
new methods include target enrichment strategies \citep{Mamanova:2010vn} like
exon capture \citep{22900609}, anchored phylogenomics \citep{Lemmon:2012fk} and
ultra-conserved elements \citep{Faircloth:2012ys}, as well as RAD sequencing
\citep{Baird:2008ly,Davey:2011zr}. As a result genome-wide samples of large
numbers of loci from multiple individuals and multiple species have become
increasingly common. This trend is rapidly shifting the \textit{modus operandi}
of systematic biology from phylogenetics to phylogenomics. This move to phylogenomics has also heralded a rapid development and uptake
of species tree inference methods that acknowledge and model the discordance
among individual gene trees. As with the field of phylogenetics, there is a
broad acceptance that probabilistic model-based methods are preferable, however
the amount of data produced by next-generation technologies has also spurred the
development of faster methods that do not utilize all the available data and
employ statistical shortcuts such as admitting no uncertainty in individual gene
trees \citep{Kubatko:2009qf, Liu:2009ve}.

\subsection{Bayesian species tree estimation}

The theory of incomplete lineage sorting and its implications for phylogenetic
inference has been appreciated for some time \citep{Pamilo:1988fk}, and early
approaches to applying this theory inferred the species tree that minimizes deep
coalescences using gene tree parsimony
\citep{Maddison01091997,Page1997231,Slowinski01101999}. The fully probabilistic
application of the theory to molecular sequence analysis has only begun more
recently with the introduction of Bayesian implementations of the multispecies
coalescent
\citep{RannalaYang2003,Edwards03042007,liu2008best,Liu:2008kx,*BEAST}. This
model embeds gene trees within a birth-death or pure Yule species tree, and
within each lineage (or branch) of the species tree, gene trees are assumed to
follow a coalescent process \citep{*BEAST}. Prior to the development of these
methods it was necessary to assume that the history of each gene is shared and
equal to the history of the species tree being studied.

However, gene trees evolve within a species tree and the
approximation of equating them becomes increasingly problematic as one samples
more loci, when in reality each have distinct gene tree topologies and
divergence times. The multispecies coalescent brings together coalescent and
birth-death models of time-trees into a single model. It describes the
probability distribution of one or more gene trees that are nested inside a
species tree. The species tree describes the relationship between the sampled
species, or sometimes, sampled populations that have been separated for long
periods of time relative to their population sizes. In the latter case it may be
referred to as a \textit{population tree} instead.

The initial implementations of the multispecies coalescent made very simple
assumptions including no recombination within each locus and free recombination
between loci. While these simple assumptions can be robust to violation,
including some forms of gene flow \citep{Heled:2013fk} (but see
\cite{Leache01012014}), researchers have begun to acknowledge that additional
processes (such as hybridization) may need to be incorporated
\citep{Joly2009,Kubatko01102009,chung2011comparing,yu2011coalescent,camargo2012accuracy}. A number of
simulation studies have also looked at various facets of performance of Bayesian
species tree estimation including the influence of missing data
\citep{Wiens:2011uq}, the influence of low rates and rate variation among loci
\citep{Lanier:2014kx} and comparisons of performance with ``supermatrix''
concatenation approaches
\citep{DeGiorgio:2010ys,Larget:2010zr,leache2011accuracy,Bayzid:2013vn}.

Although these modelling advances are exciting, in the face of a next-generation
data deluge, this study asks and answers the following, heretofore unanswered
questions: (i) How do fully Bayesian multispecies coalescent methods scale to
data sets of hundreds of loci? (ii) How much more accurate will phylogenetic
species tree estimates be with more sequence data? (iii) When should one use a
multispecies coalescent approach instead of computationally more efficient
\BayesianSupermatrix{} approaches, or summary methods which do not use all
available data? To address the first of these questions we investigate the
computational performance of the *BEAST implementation of the multispecies
coalescent \citep{*BEAST}, so as to assess the feasibility of conducting
phylogenomic analyses using existing computational tools. To shed light on the
second question we investigate how estimation accuracy improves with increasing
loci.

To address the final question, we investigate how the statistical accuracy of
the multispecies coalescent compares with concatenation across a broad range of
conditions. We also investigate the statistical accuracy of the multispecies
coalescent, supermatrix and summary methods using simulations based on two
published sequence data sets; RAD tag sequences from a study of the
Sino-Himalayan plant clade \textit{Cyathophora} \citep{Eaton01092013}, and
RNA-seq assemblies from a study of primates \citep{Perry01042012}.
\textit{Cyathophora}, a section of the genus \textit{Pedicularis} originating in
the late Miocene or the Pliocene, is probably no older than 8 Ma
\citep{YangWang2007} and is therefore a shallow study system. In contrast
primates are a deep study system, as the oldest split in this order is estimated
to have occurred in the Cretaceous around 80 Ma
\citep{Tavare2002,Steiper2006384,Wilkinson2011}.

\section{Methods}

Using simulation, we investigated the trends in computational
performance and statistical accuracy of the multispecies coalescent model as
implemented in BEAST 2 (*BEAST), and its statistical accuracy relative to other
methods of species tree inference. In designing these simulation studies there
were a number of parameters to consider. The key parameters that might
determine performance of inference under the multispecies coalescent are:

\begin{description}
	\item{\makebox[1in][r]{$\nS{}$} : The number of species.}
	\item{\makebox[1in][r]{$\nI{}$} : The number of individuals sampled per species.}
	\item{\makebox[1in][r]{$\nL{}$} : The number of independent loci.}
	\item{\makebox[1in][r]{$n_s$} : The number of sites in a single locus.}
	\item{\makebox[1in][r]{$N_e$} : The effective population sizes of extant and ancestral species.}
	\item{\makebox[1in][r]{$\tau$} : The branch lengths in units of time or expected substitutions.}
\end{description}

Of these parameters it is the number of loci $\nL{}$, the number of sites in a
single locus $n_s$, and the number of
individuals per species $\nI$ that are largely determined by experimental
design. In addition, a complete specification of a multispecies coalescent
model requires a speciation model (parameterized model of the species tree), a
substitution model (model of the relative rates and base frequencies) and a
clock model describing the absolute rate of evolution across the branches of
each gene tree. In the following sections we describe the choices of parameters,
models and simulation conditions for our computational experiments.

Species and gene trees for all experiments were simulated using biopy
(\url{http://www.cs.auckland.ac.nz/~yhel002/biopy/}), which simulates gene trees contained within species trees
according to the multispecies coalescent process. Sequence alignments were also simulated using biopy for experiment 1
and 2, and Seq-Gen \citep{Rambaut01061997} was used to simulate nucleotide
alignments for experiment 3.

\subsection{Experiment 1: Performance of *BEAST with increasing numbers of loci}

The first set of simulations we performed was primarily aimed at understanding the
effect that increasing the number of loci has on the computational performance
and statistical accuracy of Bayesian species tree estimation. We simulated 100
random (rapidly speciating) species trees of each of three different sizes,
$\nS{}=5,8,13$, using the birth-death process
\citep{Kendall1948b,Nee:1994fk,Gernhard:2008uq}. In all cases the speciation
rate $\lambda=1$ and the extinction rate $\mu=0.2$ (nominally per million
years). For 5 species trees we considered $\nI{}=2,4,8$, for 8 species trees
$\nI{}=2,4$ and for 13 species trees $\nI{}=2$. For each combination of $\nS$
and $\nI$ we simulated up to 256 gene trees. Gene alignments were simulated from these
gene trees using an HKY substitution model \citep{Hasegawa1985} and a strict clock. All sequences
were simulated with a substitution rate of 1\% per lineage per million years, a
transition/transversion ratio $\kappa$ of 4, equal base frequencies and a strict
clock. For each *BEAST analysis, the substitution rate was fixed at 1\%, and a
single $\kappa$ value and set of base frequencies for all loci was estimated.
The locus length was 200 sites each to mimic short-read next-generation sequence
data. Finally, we drew successively larger subsets of each group of
alignments to form a set of *BEAST analyses \citep{*BEAST}. We considered
increasing numbers of loci on a logarithmic scale, i.e. $\nL{} \in \{2, 4, 8,
16, 32, 64, 128, 256\}$.

If the effective sample size (ESS) of either the log posterior or the
age of the species tree in an analysis was not $\ge200$ after the initial MCMC
chain was completed, we used the \textit{resume} function in BEAST 2
\citep{BEAST2} to extend the MCMC chain from the final state of the previous
run, until sufficient samples were obtained to achieve a minimum ESS of 200. All
statistics and trees for each set of 100 replicates were logged at a sampling
rate of 1 sample per 25000 states, and the MCMC chains that needed extension
were combined into a single long chain. Pseudocode for the experimental protocol
can be found in Algorithm~S1 in supplementary information.

ESS per hour was not calculated using the total CPU time for the combined chain
because resumed runs were not restricted to a single type of CPU and hence were
not directly comparable. Instead, the initial MCMC chain for each condition and
replicate was restricted to a single type of CPU (Intel E5-2680 @ 2.70 GHz), and
million states per hour of CPU time was calculated based on the number of states
and CPU time of the initial chain. To calculate ESS per million states, the ESS
of the age of the species tree was divided by the million post-burnin states
in the combined chain. Finally to calculate ESS per hour, ESS per million states
was multiplied by million states per hour.

The main measure of error used in this study, ``relative species tree error,''
incorporates both topological and branch length error by building on the
previously described measure ``rooted branch score'' \citep[RBS;][]{24093883}.
Given two trees $T_1$ and $T_2$, the sets of monophyletic clades $c$ present in
each tree are defined as $\mathbb{C}_1$ and $\mathbb{C}_2$. The length of the
branch which extends rootward from the most recent common ancestor (MRCA) of a
clade is defined as $b(c)$. Given these definitions, the rooted branch score is
defined as the sum of all absolute differences in branch lengths $b(c)$ between
trees $T_1$ and $T_2$:

\begin{equation}
RBS(T_1, T_2) = \sum_{c \in {\mathbb{C}_1} \cup {\mathbb{C}_2}} |b^{(1)}(c) - b^{(2)}(c)|
\end{equation}

By convention, the branch length of a clade that is missing from a tree is zero,
so the topological error of absent or erroneous clades will be weighted by the
true or estimated branch length respectively.
We define the relative species tree error $e_{T}$ to be the posterior
expectation of the rooted branch score distance $RBS$ between the estimated
species tree $\hat{T}$ and the true species tree $T_{true}$, normalized by the
tree length of the true species tree $L_{true}$:

\begin{equation}
e_{T} = \frac{\frac{1}{k} \cdot \sum_{i=1}^{k} RBS(T_{true}, \hat{T}_i)}{L_{true}}
\end{equation}

This measure summarizes the error over the entire posterior distribution by
averaging the RBS for each $i$ posterior sample $\hat{T}_i$ drawn from the entire
set of posterior samples of size $k$. We normalize by the length of the true
species tree to make the error comparable between species trees of differing
units and/or number of species.

A post-hoc analysis was performed to investigate the residual variation in ESS rates and
relative species tree error, after accounting for the number of loci,
individuals and species in each replicate. Spearman's rank correlation was used
to calculate correlation coefficients between the residuals and various tree and
alignment parameters. P-values for each correlation were computed using
asymptotic \textit{t} approximation, and then corrected for multiple comparisons
based on 48 tests per set of residuals \citep{BenjaminiHochberg1995}.

Mean population size was calculated as the mean of all
per-branch effective population sizes. Species tree asymmetry is the variance
$\sigma^2_N$ in the number of nodes between each tip and the tree root
\citep{KirkpatrickSlatkin1993}. Mean tree height difference is the mean
difference in height between each gene tree and the species tree. Mean deep
coalescences is the mean number of deep coalescences for each gene as
calculated by DendroPy 4.0.3 \citep{Sukumaran15062010}. The mean parsimonious
mutations is the parsimonious (minimum) number of mutations
required per site given the true gene tree, again calculated by
DendroPy. Mean variable site count is the mean number of sites per locus with
more than one extant allele, and mutations per variable site is the total number
of parsimonious mutations required divided by the total number of variable sites.

Experiment 1 was performed using the Pan cluster provided by New Zealand
eScience Infrastructure and hosted at the University of Auckland
(\url{http://www.eresearch.auckland.ac.nz/en/centre-for-eresearch/research-facilities/computing-resources.html}).
This high performance compute cluster provides access to Linux compute nodes
with 2.7 and 2.8GHz Intel Xeon CPUs, and approximately 8GB of RAM per CPU core.

\subsection{Experiment 2: Comparing a Bayesian multispecies coalescent approach with a \BayesianSupermatrix{} approach}

In the second set of simulations we compare the statistical accuracy of the
multispecies coalescent to partitioned concatenation, both as implemented in BEAST
2. We refer to these methods as *BEAST and \BayesianSupermatrix{} respectively.
Specifically we tested the hypothesis that the comparative accuracy would depend
on mean branch length in coalescent units of $\tau(2N_e)^{-1}$.

For every combination of $\nS{}=4,5,6,8$ and $\nL{}=1,2,4$ we simulated species
trees with a range of branch lengths in coalescent units. In order to vary
branch lengths, species trees were simulated with expected root heights of
$R=\frac{1}{2},1,2,4,8,16$ (nominally in millions of years) and population sizes
chosen from $N_e=\frac{1}{4},\frac{1}{2},1$ (nominally in units of million
individuals), changing the coalescent branch length unit numerator and
denominator respectively. Additional expected root heights were included where
the most accurate method switches from *BEAST to \BayesianSupermatrix{}, to
obtain denser sampling in that part of parameter space.

Species trees were generated under the pure birth Yule model \citep{Yule1924}.
The birth rate for each combination of parameters was set to $\lambda =
\frac{1}{R} \sum_{k=2}^\nS{} \frac{1}{k}$, that is, the birth rate which generates
trees with an expected root height of $R$. These settings roughly correspond to
mammalian nuclear genes of species with an effective population size of
one-quarter, one half or one million individuals.

A single individual per species was simulated for all loci. We used the
Jukes-Cantor substitution model \citep{Jukes1969} and a strict clock model for
each locus, but with rate variation between loci. The mutation rate for the
first locus was fixed at $\mu_0=0.01$, and the rates for other loci drawn from
the range $[\mu_0 / F, \mu_0 \times F]$. We used $F=3$, giving a factor of 9
between the fastest and slowest possible rates. The rate was drawn in log space,
so there is equal density of slower and faster rates around $\mu_0$. The number
of sites per alignment ($n_s$) was fixed at 1000.

We generated 100 replicates for each combination of $\nS{}$, $\nL{}$, $R$ and
$N_e$. For each unique combination of $\nS{}$, $R$ and $N_e$ only one set of 100
species trees was generated and used (regardless of $\nL{}$) to minimize species
tree sampling error when analyzing the effect of increasing $\nL{}$. Gene trees
and extant sequences were generated separately for each replicate and for each
value of $\nL{}$.

Both \BayesianSupermatrix{} and *BEAST analyses used a Yule prior on the
species tree, with a uniform prior of $[{^1/_{100}},100]$ on $\lambda$, and
$\nL{}$ partitions with a strict clock model for each, where the clock rate for
the first partition is fixed to the truth ($\mu_0$) and the other rates were
estimated. The *BEAST effective population size hyperparameter (popMean) was
given a uniform prior in the range $[\frac{1}{5},5]$, and all population sizes
were estimated.

The \BayesianSupermatrix{} analysis used a fixed chain length of 4 million
states, sampling every 1000 states. The *BEAST analysis used a fixed chain
length of 40 million states, sampling every 10,000 states. The ESS values of the
posterior, likelihood and prior statistics of each chain were estimated, and
replicates where the ESS was $<200$ for any of those statistics were discarded.
For each combination of $\nS{}$, $\nL{}$ and method there were never more
than 4\% of replicates discarded for this reason (Figure~S10). As with
experiment 1, this experiment was performed using the NeSI Pan cluster.

\subsection{Experiment 3: Many-method comparison of species tree inference using parameters estimated from two phylogenomic data sets}

The purpose of the third set of simulations was two-fold: to check that the
trends in statistical accuracy observed for the first two sets of simulations
held for empirically derived simulations, and to compare statistical accuracy
across a range of species tree inference methods.  To simulate more realistic
trees and sequences, we derived a range of properties and phylogenetic
parameters from two empirical phylogenomic data sets for use as simulation
parameters.

The biallelic species tree inference method SNAPP \citep{Bryant01082012} was
used to estimate speciation birth rates and effective population sizes because
it did not require phasing the sequence data. To estimate base frequencies,
substitution rates, between-site rate variation and between-locus rate variation
we used a \BayesianSupermatrix{} analysis with a Yule prior on the
species tree. A detailed description of sequence data processing and SNAPP and
BEAST settings is given in supplementary information.

We simulated 100 replicates each of ``deep'' and ``shallow'' Yule species trees
of $\nS{}=12$ and $\nS{}=8$ respectively, using the inferred empirical birth
rates, with per-branch population sizes picked from a gamma distribution of
shape 2 and a mean equal to the mean inferred population sizes. For the deep
species trees we simulated 512 gene trees, and for the shallow species trees we
simulated 4096 gene trees within each species tree, each with two
individuals per species.

For each simulated gene tree we chose a strict clock rate from the gamma
distribution defined by the inferred shape parameters and scale parameters.
Nucleotide sequences were simulated for every locus using the empirically
derived GTR+G base frequencies, substitution rates and gamma rate variation from
the applicable study. As the shallow study used 64nt RAD tags, we picked that
fixed length for sequence simulations based on that study. For simulations
based on the deep study, each simulated alignment length was randomly sampled
(with replacement) from the original alignment lengths of the deep study.

Species trees were reconstructed from simulated sequences using five different
multi-locus inference methods; *BEAST, \BayesianSupermatrix{}, \MPEST{}
\citep{20937096}, RAxML version 8 \citep{Stamatakis01052014} and BIONJ
\citep{Gascuel01071997}. We tested *BEAST performance given $\nL{}=1,2,4,8$ for
the deep study based simulations and $\nL{}=1,2,4,8,16,32$ for the shallow study
based simulations. For all simulations, we tested the performance of
\BayesianSupermatrix{} given $\nL{}=1,2,4,8,16,32,64,128,256,512$. For the deep
study simulations we tested RAxML, BIONJ and \MPEST{} with
$\nL{}=1,2,4,8,16,32,64,128,512$. For the shallow study simulations we also
analyzed $\nL{}=1024,2048,4096$. Both *BEAST and \MPEST{} can infer species trees
utilizing more than one individual per species, and we tested both methods using
$\nI{}=1,2$.

All GTR+G rates were estimated for *BEAST and \BayesianSupermatrix{} analyses.
For RAxML analyses, only GTR+G substitution rates were estimated and empirical
base frequencies were used. Clock rate distribution parameters and clock rates
for each locus were estimated for *BEAST and \BayesianSupermatrix{} analyses.
Partitioning was not used (so per-locus clock rates could not be estimated) for RAxML
analyses. The RAxML maximum likelihood algorithm used was ``new rapid
hillclimbing''. Pairwise distances matrices calculated by RAxML were used to
generate neighbor-joining trees using the BIONJ algorithm implemented in PAUP*
version 4.0a142 (\url{http://paup.csit.fsu.edu/}). *BEAST and BEAST trees are implicitly rooted
because they are ultrametric, and RAxML and BIONJ trees were midpoint rooted.

\MPEST{} uses gene trees as input data, which were inferred using RAxML. The same
settings used for RAxML species tree inference were used for gene tree
inference, and gene trees were midpoint rooted. For each replicate \MPEST{}
was set to make 10 independent runs, and the species tree with the highest
pseudo-likelihood was retained for further analysis.

The BEAST and *BEAST chains were run on the Raijin cluster provided by the
National Computational Infrastructure
(\url{http://nci.org.au/systems-services/national-facility/peak-system/raijin/}).
This cluster provides access to Linux compute nodes with 2.6GHz Intel Xeon Sandy
Bridge CPUs, and 4GB of RAM was requested per run. Further details of BEAST and
*BEAST chains are provided in supplementary information. RAxML and \MPEST{} were
run on the cluster provided by the Genome Discovery Unit of the Australian
Cancer Research Foundation Biomolecular Resource Facility. Jobs on this cluster
ran on Linux compute nodes with a variety of Intel Xeon and AMD Opteron CPUs,
and 2GB of RAM was requested per RAxML or \MPEST{} job.

\clearpage

\section{Results}

\subsection{Experiment 1: Performance of *BEAST with increasing numbers of loci}

\subsubsection{Computational performance} 
\begin{figure}[htb!]
\centering
\includegraphics[height=14cm]{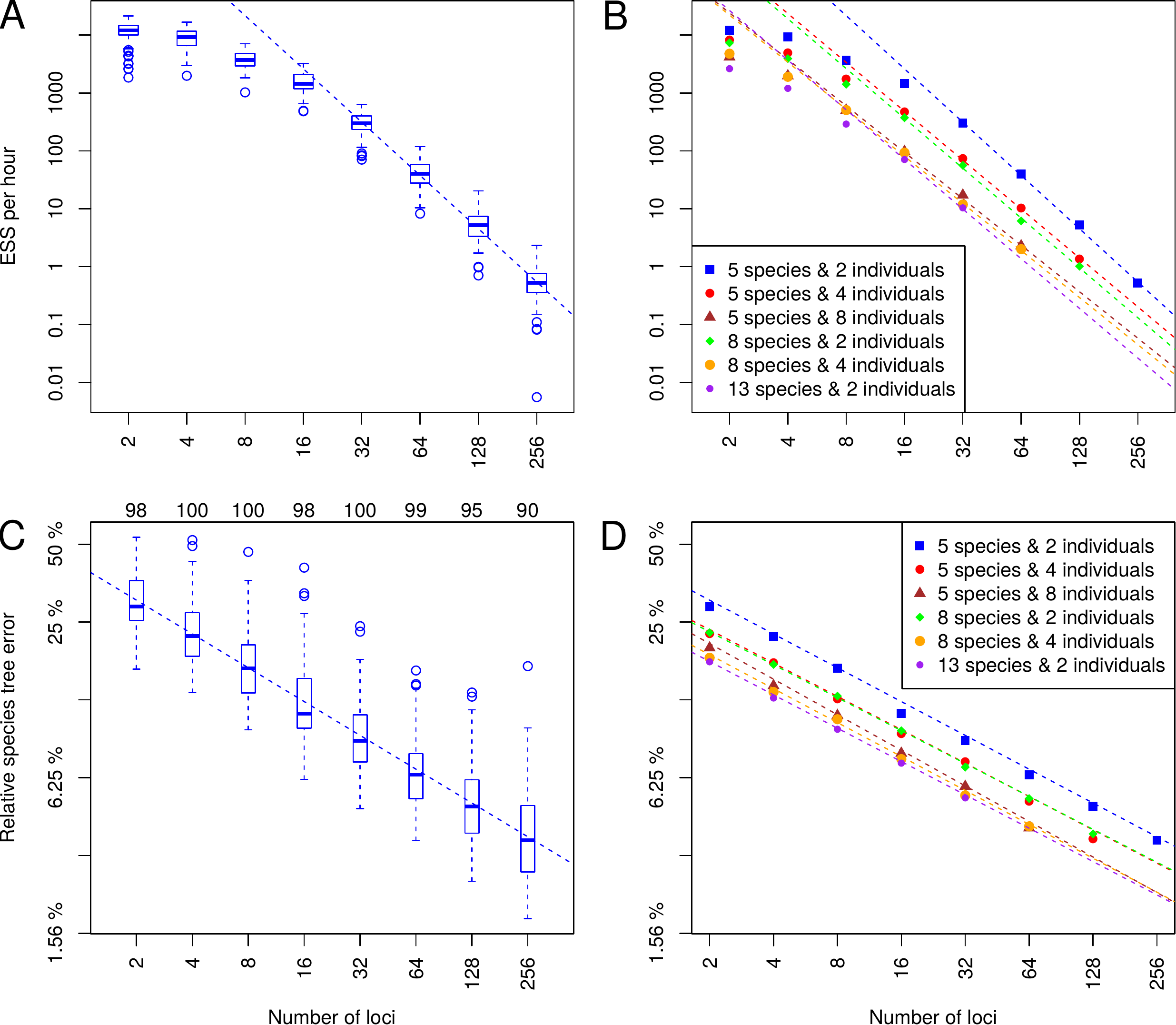}
\caption
{Trends in ESS per hour and relative species tree error as a function of the
number of loci. (a) ESS per hour for analyses of 5 species each with 2
individuals. Each box-and-whisker shows the variance in mixing across a hundred
replicate data sets for each number of loci. (b) The median ESS per hour as a
function of number of loci, with trend lines for each combination of number of
species and individuals per species. Solid shapes indicate the median value for
each category, and regression lines were calculated using all replicates for
each category. (c) Relative error for 5 species each with 2 individuals, with
each box-and-whisker showing the variance in relative error between replicates.
Numbers above the graph area indicate how many replicates were included for each number
of loci. (d) The relative error in the estimated species tree as a function of
the number of loci, with trend lines for each combination of number of species
and individuals per species. Solid shapes indicate the median value for each
category, and regression lines were calculated using all replicates for each
category with sufficient ESS.}
\label{fig:powerLawResults}
\end{figure}

We evaluated the scaling of
computational performance of *BEAST as a function of the number of loci
analyzed. We recorded the elapsed computational time for each replicate analysis
running in a single thread. This was then used to calculate the effective number
of samples per hour (ESS per hour), to measure the computational effort 
required to produce a sample from the posterior for a given number of loci. The
ESS per hour relationship (Figure~\ref{fig:powerLawResults}a,S3) suggests that a
power law fits the scaling of computational performance. The linear relationship
in the log-log plot indicates that a power law fits well for the range from 32
to 256 loci. We extrapolate that for $\nS{} = 5$, $\nI{} = 2$ and $\nL{} \geq
32$, ESS per hour follows a power law with an slope and intercept of $-3.06 \pm
0.04$ and $16.34 \pm 0.18$ respectively.

Applying this functional relationship, we could estimate the computational cost
to analyze a similar data set with a larger number of loci. For example, given 5
species and 2 individuals in the simulation, the predicted ESS per hour is 0.54
for 256 genes, which indicates it would take approximately 369 CPU hours to
obtain an ESS of 200. We can therefore estimate that a similar analysis of 1024
loci would take roughly 1064 CPU days. Nevertheless an analysis this size might
be achieved within two months by parallelizing the problem into 20 independent
MCMC chains for two months each and discarding a few days of burnin from each of
them, to achieve on the order of ten independent samples from each chain.

The slope of the median computational performance as a function of number of
loci does not vary with the number of species or the number of individuals
(Figure~\ref{fig:powerLawResults}b), although a larger range of $\nS$ and $\nI$ would
need to be examined to understand the scaling relationship of computational
performance with those quantities. For analyses larger than 5 species and 2
individuals, the power law range appears to begin at $\nL{} \geq 16$. Combining
all simulation results, a multiple linear regression describing a response variable
$Y$ (e.g. ESS per hour) as a function of three explanatory variables: number of
loci $\nL$, number of species $\nS$, and number of individuals per species
$\nI$, can be constructed as follows:

\begin{equation}
\label{eq:multivariate-model}
\log(Y) = \beta_1 log(\nL) + \beta_2 \nS + \beta_3 \nI + \alpha
\end{equation}

Taking the ESS per hour as the response variable, the linear regression
estimates of the coefficients are $\beta_1 = -2.81 \pm 0.02$, $\beta_2 = -0.42
\pm 0.01$, $\beta_3 = -0.46 \pm 0.01$, and the intercept is $\alpha = 17.98 \pm
0.13$. At least within the range of parameters examined here, it appears
that the $\beta_1$ coefficient is not greatly influenced by $\nS$ and $\nI$
(Figure~\ref{fig:powerLawResults}b).

We also considered the scaling of the number of effective samples per million
states (ESS per million states) in the MCMC analyses. This quantity is
complementary to our first result; it is easier to investigate as it does not
require running all simulations on identical and dedicated hardware.
Computational time for methods like *BEAST is dominated by the phylogenetic
likelihood, which is calculated for all site patterns given a proposed tree
\citep{Yang01031994}. Because *BEAST infers a separate gene tree for each locus,
the time per state will be linear with the number of loci assuming the average
number of site patterns per locus is independent of the total number of loci.
This assumption of independence holds for experiment 1 because loci were
subsetted uniformly.

Adapting the terminology of Equation~\ref{eq:multivariate-model}, the slope of
ESS per hour ($\beta_{1h}$) will be simply related to the slope of ESS per
million states ($\beta_{1s}$): $\beta_{1h} = \beta_{1s} + 1$. However because
CPU time per site pattern depends on the specific hardware employed, the
intercept of ESS per hour ($\alpha_h$) cannot be predicted from that of ESS per
million states ($\alpha_s$).

As expected, ESS per million states also exhibits a power law in the number of
loci (Figure~S4). By assigning the ESS per million states to $Y$ in the multiple
linear regression in Equation~\ref{eq:multivariate-model}, the estimated
coefficients are $\beta_1 = -1.87 \pm 0.02$, $\beta_2 = -0.28 \pm 0.01$,
$\beta_3 = -0.24 \pm 0.01$, and the estimated intercept is $\alpha = 9.07 \pm
0.12$. The difference in slope between ESS per million states and ESS per hour
is $(-1.87) - (-2.81) = 0.94$, very close to 1 as predicted. As with ESS per
hour, observations used for the linear regression were restricted to
$\nL{}\geq32$ for the 5 species, 2 individual case and $\nL{}\geq16$ for other
cases.

Using the example of 5 species and 2 individuals, the slope and intercept are
$-1.97 \pm 0.04$ and $7.86 \pm 0.18$ respectively, so the predicted ESS per
million states for 256 individuals is 0.047 (Figure~S4a). It would therefore
take approximately 4.3 billion states to obtain an ESS of 200. We can
extrapolate that a similar analysis of 1024 loci would require an MCMC chain of
roughly $4.3 \times \left(\frac{1024}{256}\right)^{1.97} \approx 66$ billion
states.

\subsubsection{Statistical accuracy}
We also calculated the relative error in the species tree estimate for each
replicate. For some larger analyses it was challenging to achieve
acceptable ESS values for every replicate, even with chain lengths of
several billion states and access to high performance computational
infrastructure. To retain the larger analyses without biasing statistical
accuracy, we excluded replicates in which the ESS of either the log posterior or
the species tree age was smaller than 200. All remaining replicates were used
for a linear regression analysis of the contribution of the number of loci to
relative species tree error. This analysis revealed a power law
relationship from 2 to 256 loci (Figure~\ref{fig:powerLawResults}c,S5). Given
5 species and 2 individuals, the slope
and intercept are $-0.435 \pm 0.007$ and $-0.889 \pm 0.026$ respectively, so the
relative species tree error predicted by the power law for 256 loci is 0.037.
By extrapolation we would therefore estimate that the relative error of a 1024
loci analysis would decrease to $0.037 \times
\left(\frac{1024}{256}\right)^{-0.435} \approx 0.020$.

A linear regression analysis of relative species tree error for all combinations
of $\nS{}$ and $\nL{}$ showed little variation in the trend line slope between
conditions (Figure~\ref{fig:powerLawResults}d). By assigning the relative species
tree error to $Y$ in the multiple linear regression in
Equation~\ref{eq:multivariate-model}, the estimated coefficients are
$\beta_1 = -0.433 \pm 0.003$, $\beta_2 = -0.066 \pm 0.002$, $\beta_3 = -0.070
\pm 0.002$, and the estimated intercept is $\alpha = -0.481 \pm 0.022$. More details for
all multiple linear regression models are available in supplementary
information. Trends in topology-only accuracy inferred using rooted
Robinson-Foulds (rRF) scores are also presented in supplementary information
(Figure~S9, Table~S12).

Finally, we also analyzed the number of species tree topologies sampled in each
posterior distribution. It appears that for the analyses involving 8
and 13 species there is a rapid reduction in the number of topologies in the
95\% credible set with increasing numbers of loci, but it does not
follow a power law (Figure~S7).

\subsubsection{Post-hoc analysis of convergence and species tree error}
Experiment 1 was designed to investigate the relationship between the number of
loci $\nL{}$, number of species $\nS{}$ and number of individuals $\nI{}$ on ESS
rates and statistical accuracy. While these variables explained most of the
variation in ESS rates and accuracy, residual variation was present
between the 100 replicates of each combination of $\nL{}$, $\nS{}$ and $\nI{}$
(Figure~\ref{fig:powerLawResults}a,c). The correlations between this residual
variation and a collection of phylogenetic statistics that could be extracted from
the simulated trees and alignments were studied in a post-hoc analysis.

\vspace{0.25in}
\begin{table}[htb!]
\centering
\caption{Spearman correlation of tree and alignment parameters with ESS per hour.}
\label{tab:ess-cor}
\begin{threeparttable}
\begin{adjustbox}{center}
\small
$\begin{array}{|c|l|l|l|l|l|l|}
\multicolumn{1}{c}{} & \multicolumn{1}{c}{5\nS{}, 2\nI{}} & \multicolumn{1}{c}{5\nS{}, 4\nI{}} & \multicolumn{1}{c}{5\nS{}, 8\nI{}} & \multicolumn{1}{c}{8\nS{}, 2\nI{}} & \multicolumn{1}{c}{8\nS{}, 4\nI{}} & \multicolumn{1}{c}{13\nS{}, 2\nI{}}\tabularnewline
\hline
\text{Species tree height} & \hphantom{-}0.068 & \hphantom{-}0.222^{***} & \hphantom{-}0.362^{***} & -0.036 & \hphantom{-}0.180^{***} & \hphantom{-}0.120\tabularnewline
\hline
\text{Mean population size} & \hphantom{-}0.075 & -0.048 & -0.086 & -0.020 & -0.101 & \hphantom{-}0.121\tabularnewline
\hline
\text{Species tree asymmetry} & -0.238^{***} & -0.088 & -0.045 & -0.125^{*} & \hphantom{-}0.013 & -0.068\tabularnewline
\hline
\text{Mean deep coalescences} & -0.122^{**} & -0.225^{***} & -0.295^{***} & \hphantom{-}0.020 & -0.079 & \hphantom{-}0.044\tabularnewline
\hline
\text{Mean parsimonious mutations} & \hphantom{-}0.099 & \hphantom{-}0.148^{***} & \hphantom{-}0.122^{*} & -0.013 & \hphantom{-}0.124^{*} & \hphantom{-}0.074\tabularnewline
\hline
\text{Mean variable site count} & \hphantom{-}0.088 & \hphantom{-}0.228^{***} & \hphantom{-}0.294^{***} & -0.045 & \hphantom{-}0.146^{**} & \hphantom{-}0.042\tabularnewline
\hline
\text{Mean tree height difference} & \hphantom{-}0.246^{***} & \hphantom{-}0.355^{***} & \hphantom{-}0.315^{***} & \hphantom{-}0.421^{***} & \hphantom{-}0.340^{***} & \hphantom{-}0.398^{***}\tabularnewline
\hline
\text{Mutations per variable site} & \hphantom{-}0.030 & -0.066 & -0.123^{*} & \hphantom{-}0.046 & \hphantom{-}0.016 & \hphantom{-}0.057\tabularnewline
\hline
\end{array}$
\end{adjustbox}
\begin{tablenotes}
\small
\item {*}: $p < 0.05$, {**}: $p < 0.01$, {***}: $p < 0.001$.
\end{tablenotes}
\end{threeparttable}
\end{table}

The only tree or alignment statistic that was significantly correlated with ESS
per hour consistently across all conditions was mean
tree height difference (Table~\ref{tab:ess-cor}). This statistic is the mean
difference in height between each gene tree and the species tree. The positive
correlation observed for this parameter suggests that when gene trees are taller
relative to the species tree, the ESS rate will be higher and *BEAST will
converge more quickly.

\vspace{0.25in}
\begin{table}[htb!]
\centering
\caption{Spearman correlation of tree and alignment parameters with species tree error.}
\label{tab:error-cor}
\begin{threeparttable}
\begin{adjustbox}{center}
\small
$\begin{array}{|c|l|l|l|l|l|l|}
\multicolumn{1}{c}{} & \multicolumn{1}{c}{5\nS{}, 2\nI{}} & \multicolumn{1}{c}{5\nS{}, 4\nI{}} & \multicolumn{1}{c}{5\nS{}, 8\nI{}} & \multicolumn{1}{c}{8\nS{}, 2\nI{}} & \multicolumn{1}{c}{8\nS{}, 4\nI{}} & \multicolumn{1}{c}{13\nS{}, 2\nI{}}\tabularnewline
\hline
\text{Species tree height} & -0.734^{***} & -0.582^{***} & -0.330^{***} & -0.702^{***} & -0.537^{***} & -0.580^{***}\tabularnewline
\hline
\text{Mean population size} & \hphantom{-}0.103^{*} & \hphantom{-}0.078 & \hphantom{-}0.006 & \hphantom{-}0.118^{*} & \hphantom{-}0.004 & \hphantom{-}0.076\tabularnewline
\hline
\text{Species tree asymmetry} & \hphantom{-}0.041 & \hphantom{-}0.011 & \hphantom{-}0.035 & -0.170^{***} & -0.181^{***} & -0.050\tabularnewline
\hline
\text{Mean deep coalescences} & \hphantom{-}0.665^{***} & \hphantom{-}0.573^{***} & \hphantom{-}0.273^{***} & \hphantom{-}0.647^{***} & \hphantom{-}0.522^{***} & \hphantom{-}0.591^{***}\tabularnewline
\hline
\text{Mean parsimonious mutations} & -0.387^{***} & -0.199^{***} & -0.025 & -0.372^{***} & -0.184^{***} & -0.378^{***}\tabularnewline
\hline
\text{Mean variable site count} & -0.587^{***} & -0.494^{***} & -0.242^{***} & -0.607^{***} & -0.530^{***} & -0.642^{***}\tabularnewline
\hline
\text{Mean tree height difference} & \hphantom{-}0.194^{***} & \hphantom{-}0.186^{***} & \hphantom{-}0.196^{***} & \hphantom{-}0.173^{***} & \hphantom{-}0.207^{***} & \hphantom{-}0.127^{*}\tabularnewline
\hline
\text{Mutations per variable site} & \hphantom{-}0.416^{***} & \hphantom{-}0.306^{***} & \hphantom{-}0.152^{**} & \hphantom{-}0.333^{***} & \hphantom{-}0.220^{***} & \hphantom{-}0.148^{*}\tabularnewline
\hline
\end{array}$
\end{adjustbox}
\begin{tablenotes}
\small
\item {*}: $p < 0.05$, {**}: $p < 0.01$, {***}: $p < 0.001$.
\end{tablenotes}
\end{threeparttable}
\end{table}

In contrast to ESS per hour, several statistics were consistently significantly
correlated with relative species tree error (Table~\ref{tab:error-cor}). The
height of the species tree and the number of variable sites per locus were
negatively correlated with relative error. This result is somewhat intuitive, as
taller species trees will have longer branches which are easier to resolve, and
the number of variable sites is an obvious proxy for the amount of information
in each locus. Relative error was positively correlated with the mean number of
deep coalescences and the number of mutations per variable site. Those
correlations suggest that data sets with more incomplete lineage sorting will be
more difficult to resolve, and that saturated sites may increase uncertainty.

\subsection {Experiment 2: Statistical accuracy of \BayesianMultispeciesCoalescent{} relative to \BayesianSupermatrix{}}

To assess the statistical accuracy of the \BayesianMultispeciesCoalescent{}
relative to the standard \BayesianSupermatrix{} approach, we conducted a
simulation study where we simulated species trees with a broad range of mean
branch lengths for varying numbers of species and loci. Gene coalescences occur
prior to species divergence times, and the severity of this discrepancy will
depend on species tree branch lengths in units of coalescent time. Because the
multispecies coalescent accounts for this phenomenon but the
\BayesianSupermatrix{} approach does not, we expected the multispecies
coalescent to outperform the \BayesianSupermatrix{} approach for trees with
shorter branch lengths.

The ``species tree error ratio''
$\nicefrac{e_{T_a}}{e_{T_b}}$ is a measure of the comparative accuracy and is specified as follows, where $a$ is \BayesianMultispeciesCoalescent{} and $b$
is \BayesianSupermatrix{}:

\begin{equation}
\frac{e_{T_a}}{e_{T_b}} = \frac{\frac{1}{k_a} \cdot \sum_{i=1}^{k_a} RBS(T_{true}, \hat{T}_{ai})}
{\frac{1}{k_b} \cdot \sum_{i=1}^{k_b} RBS(T_{true}, \hat{T}_{bi})}
\end{equation}

\begin{figure}[htb!]
\centering
\includegraphics[height=16cm]{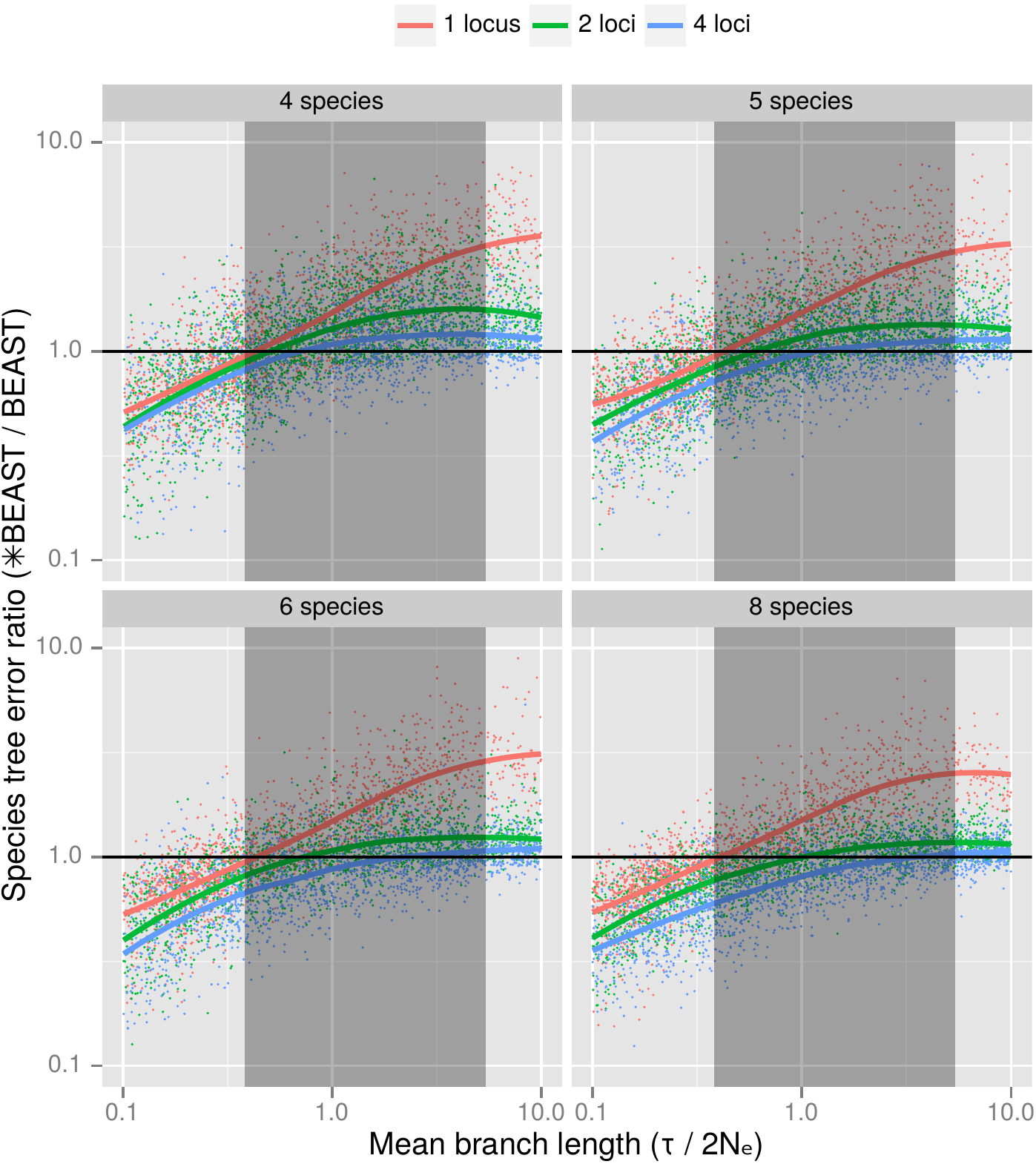}
\caption
{Species tree error ratio (*BEAST/BEAST) as a function of the average species
tree branch length (in coalescent units) for trees of 4, 5, 6 and 8 species.
Data points are below 1 (black line) where the *BEAST error is lower than the
BEAST error, indicating that *BEAST was more accurate than BEAST. Data points
above 1 show the opposite. Only results with both mean branch lengths and error
ratios between 0.1 and 10.0 are included. The red, green and blue lines show the
local regression for one, two and four locus estimates respectively. The shaded
region indicates where the crossover point depended on the combination of
simulation parameters chosen, *BEAST was always preferred for average branch
lengths shorter than this zone.}
\label{fig:branchScoreRatio}
\end{figure}

Values below 1 indicate lower error, or equivalently superior accuracy, when
using \BayesianMultispeciesCoalescent{} instead of \BayesianSupermatrix{}. For
all numbers of species tested, the statistical accuracy of
\BayesianMultispeciesCoalescent{} was superior to \BayesianSupermatrix{} for
trees with shorter mean branch lengths (Figure~\ref{fig:branchScoreRatio}).
Using LOESS regression, it is clear that as the number of loci increases,
\BayesianMultispeciesCoalescent{} performance improves relative to
\BayesianSupermatrix{} because for a given mean branch length, the species tree
error ratio decreases as the number of loci increases
(Figure~\ref{fig:branchScoreRatio}).

For all numbers of species and loci tested, there is a mean branch length
crossover point where for shorter mean branch lengths,
\BayesianMultispeciesCoalescent{} is expected to outperform
\BayesianSupermatrix{}, and \textit{vice versa} for longer mean branch lengths.
The crossover point depends on the number of loci; as the number of loci
increases, the point shifts right (Figure~\ref{fig:branchScoreRatio}),
indicating that \BayesianMultispeciesCoalescent{} is expected to outperform
\BayesianSupermatrix{} for a larger range of mean branch lengths, consistent
with the general trend of improved performance of
\BayesianMultispeciesCoalescent{} when increasing the number of loci.

Within the parameter region explored in this experiment, depending on the number
of species, loci and the effective population sizes, the crossover point was
found in the range $0.382\tau(2N_e)^{-1}$ to $5.416\tau(2N_e)^{-1}$
(Figure~S11). For mean branch lengths shorter than $0.382\tau(2N_e)^{-1}$,
\BayesianMultispeciesCoalescent{} was preferred regardless of the parameters
explored, even when using a single locus (Figure~\ref{fig:branchScoreRatio}).
The crossover point given a single locus was always below
$0.5\tau(2N_e)^{-1}$ (Figure~S11) and given longer mean branch lengths the
relative performance of \BayesianSupermatrix{} was higher than for
multi-locus inference (Figure~\ref{fig:branchScoreRatio}). This implies that
\BayesianMultispeciesCoalescent{} is still useful for single locus studies of
species trees with short branches, but should be applied with caution.

\subsection {Experiment 3: Inferred parameters of phylogenomic data sets and multi-method comparison}

Sequence data sets from two published studies were realigned and reanalyzed to
calculate their empirical properties and phylogenetic parameters. Besides the
expected difference in speciation rate (which for the shallow study rate was over six
times faster, corresponding to much shorter branch lengths), the shallow plant
study sequences were very AT rich, whereas the deep primate study sequences were
moderately GC rich (Table~\ref{tab:real-data-stats}). $C \leftrightarrows T$
substitutions were a greater proportion of all substitutions for the deep study,
but the between-site gamma rate variation was flatter. The mean effective
population size $N_e$ of the deep study was estimated to be only 2.4\% that of the
shallow study.

\begin{table}[htb!]
\centering
\caption{Experiment 3 data set properties and mean values of inferred parameters.}
\label{tab:real-data-stats}
\begin{threeparttable}
\begin{tabular}{|c||c|c|}
\hline
Phylogenetic depth & Shallow & Deep\tabularnewline
\hline
Clade name & Cyathophora & Primates\tabularnewline
\hline
Taxonomic rank & Section & Order\tabularnewline
\hline
Sequence data & RAD tag & RNA-seq\tabularnewline
\hline
In-group $nS{}$ & 8 & 12\tabularnewline
\hline
Base frequency: A & 0.290 & 0.266\tabularnewline
\hline
Base frequency: C & 0.212 & 0.240\tabularnewline
\hline
Base frequency: G & 0.204 & 0.263\tabularnewline
\hline
Base frequency: T & 0.294 & 0.231\tabularnewline
\hline
$A \leftrightarrows C$ rate & 0.367 & 0.152\tabularnewline
\hline
$A \leftrightarrows G$ rate & 0.940 & 0.694\tabularnewline
\hline
$A \leftrightarrows T$ rate & 0.246 & 0.100\tabularnewline
\hline
$C \leftrightarrows G$ rate & 0.305 & 0.155\tabularnewline
\hline
$C \leftrightarrows T$ rate & 1.000 & 1.000\tabularnewline
\hline
$G \leftrightarrows T$ rate & 0.353 & 0.127\tabularnewline
\hline
Gamma rate variation & 0.0383 & 0.233\tabularnewline
\hline
Speciation birth rate & 125.3 & 20.7\tabularnewline
\hline
Per-branch $N_e$ & $6.35 \times 10^{-3}$ & $1.53 \times 10^{-4}$\tabularnewline
\hline
Locus length & 64nt & 110--3511nt\tabularnewline
\hline
Clock variation shape & 6.22 & 5.15\tabularnewline
\hline
Clock variation scale & 0.173 & 0.195\tabularnewline
\hline
\end{tabular}
\begin{tablenotes}
\small
\item All inferred parameters are rounded to three significant figures or one
decimal place, whichever is more precise.
\end{tablenotes}
\end{threeparttable}
\end{table}

The original publication of \textit{Cyathophora} sequences and phylogeny
suggested that \textit{P. rex} subsp. \textit{rockii} is sister to subsp.
\textit{rex} and subsp. \textit{lipskyana} \citep{Eaton01092013}. The most
common species tree topology seen in both SNAPP and \BayesianSupermatrix{}
posterior distributions supports this placement (Figure~S16,S17). The original
study left open the question of \textit{P. thamnophila} monophyly but raised the
possibility that the apparent paraphyly of this species, as replicated by our
reanalysis, is an artifact of introgression \citep{Eaton01092013}. Species trees
inferred by SNAPP and \BayesianSupermatrix{} from reanalysis of the deep
phylogenetic study (Figure~S18,S19) agreed with the accepted primate phylogeny
\citep{Perry01042012}.

\subsubsection{Analysis of empirical-based simulations}
We simulated species trees, gene trees and sequences based on the
estimated parameters of both data sets (Table~\ref{tab:real-data-stats}), and
refer to these simulations as shallow and deep phylogenetic simulations
respectively. The mean branch length of the simulated shallow species trees was
$0.539\tau(2N_e)^{-1}$, compared to $159.8\tau(2N_e)^{-1}$ for the simulated
deep species trees. We computed the relative species tree error for all *BEAST
analyses of these simulations.

\begin{figure}[htb!]
\centering
\includegraphics[height=10cm]{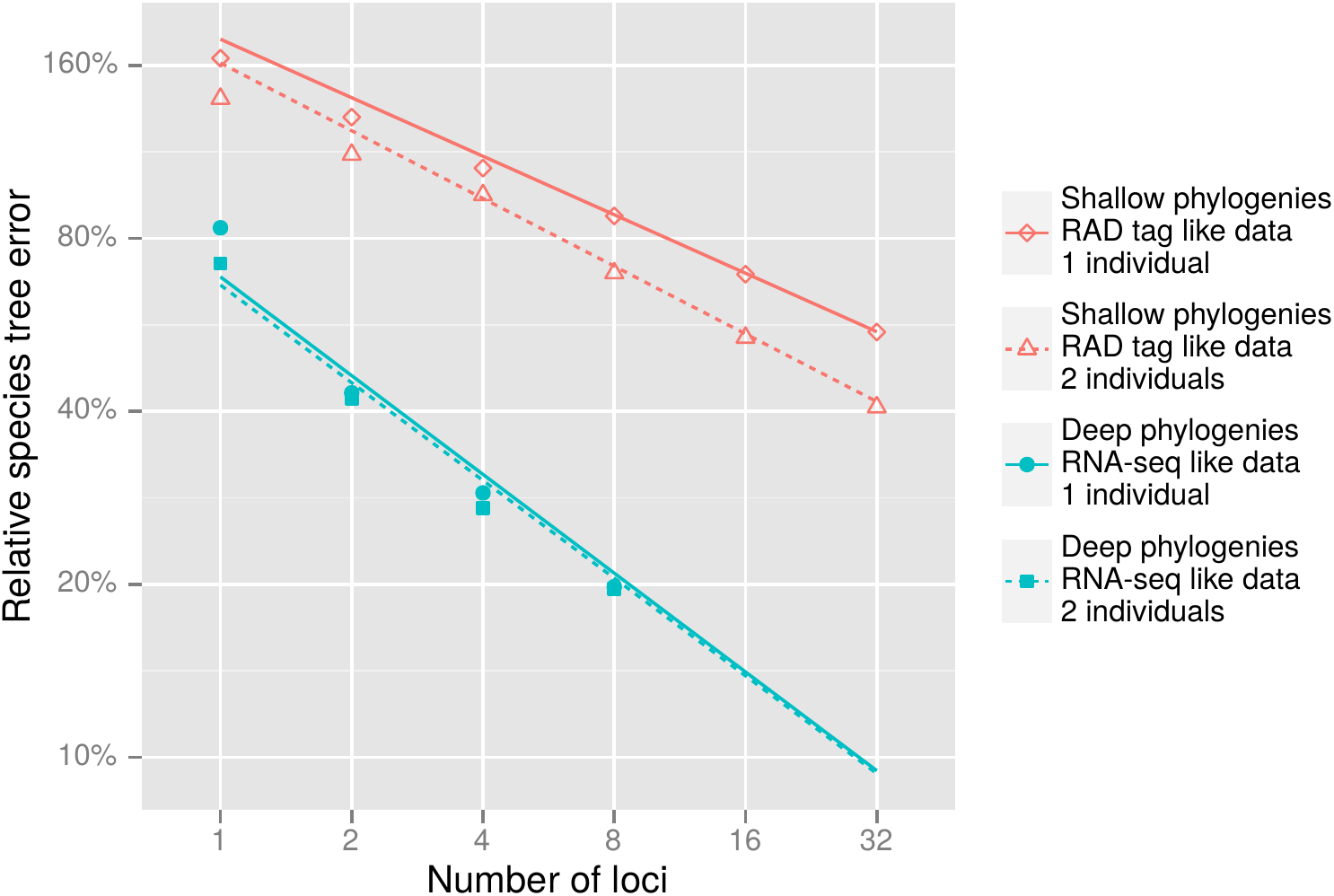}
\caption
{The relative species tree error as a function of the number of loci for
empirical-based simulations. Both shallow and deep phylogenetic simulation
results are presented. Solid and hollow shapes are the median value for each
category, and regression lines were calculated using all replicates for each
category.}
\label{fig:realisticPowerLaws}
\end{figure}

The relative species tree errors for all values of $\nL{}$ and $\nI$ considered
were computed for both simulation types. A
power law appeared to fit the relationship between relative error and number of
loci for values of $\nL{}\geq2$, so log-log linear regression analyses were
restricted to $\nL{}\geq2$. The log-log slope connecting relative error
and the number of loci appears mostly independent of $\nI{}$ for shallow
phylogenetic simulations. For deep
simulations, the trend lines for $\nI{}=1$ and $\nI{}=2$ were very close,
implying that multiple individuals did not improve accuracy for those
simulations (Figure~\ref{fig:realisticPowerLaws}).

This result is consistent with the initial set of
simulations reported in ``Statistical accuracy''. However, the log-log slopes
varied substantially between *BEAST inference of shallow and deep phylogenetic
simulations. The difference in power law exponents inferred using multiple 
linear regression (Table~S13,S14) between shallow and deep
simulations was $(-0.365) - (-0.568) = 0.203$.

Results from the initial simulation study, detailed in ``Computational
performance,'' suggest that a power law relationship of ESS and number of loci
only applies to *BEAST analyses of 16 to 32 loci and above. As we only inferred
deep phylogenetic trees utilizing up to 8 loci and shallow phylogenetic trees up
to 32 loci using *BEAST, we cannot make firm conclusions regarding the scaling
laws of ESS performance using this set of simulations.

\subsubsection{Alternative methods for multi-locus phylogenetic inference} The
second analysis we conducted based on the empirically-derived shallow and deep
phylogenetic simulations was a comparison of common multi-locus methods of
species tree inference. This encompassed the Bayesian multispecies coalescent
(*BEAST), \BayesianSupermatrix{} (BEAST), \MLSupermatrix{} (RAxML),
neighbor-joining (BIONJ) and summary coalescent (\MPEST{}) methods. As some
methods provide only a single best tree estimate in place of a posterior
distribution of trees, we used common ancestor summary trees
\citep[CAT;][]{24093883} for *BEAST and \BayesianSupermatrix{} analyses in this
comparison.

\begin{figure}[htb!]
\centering
\includegraphics[height=18cm]{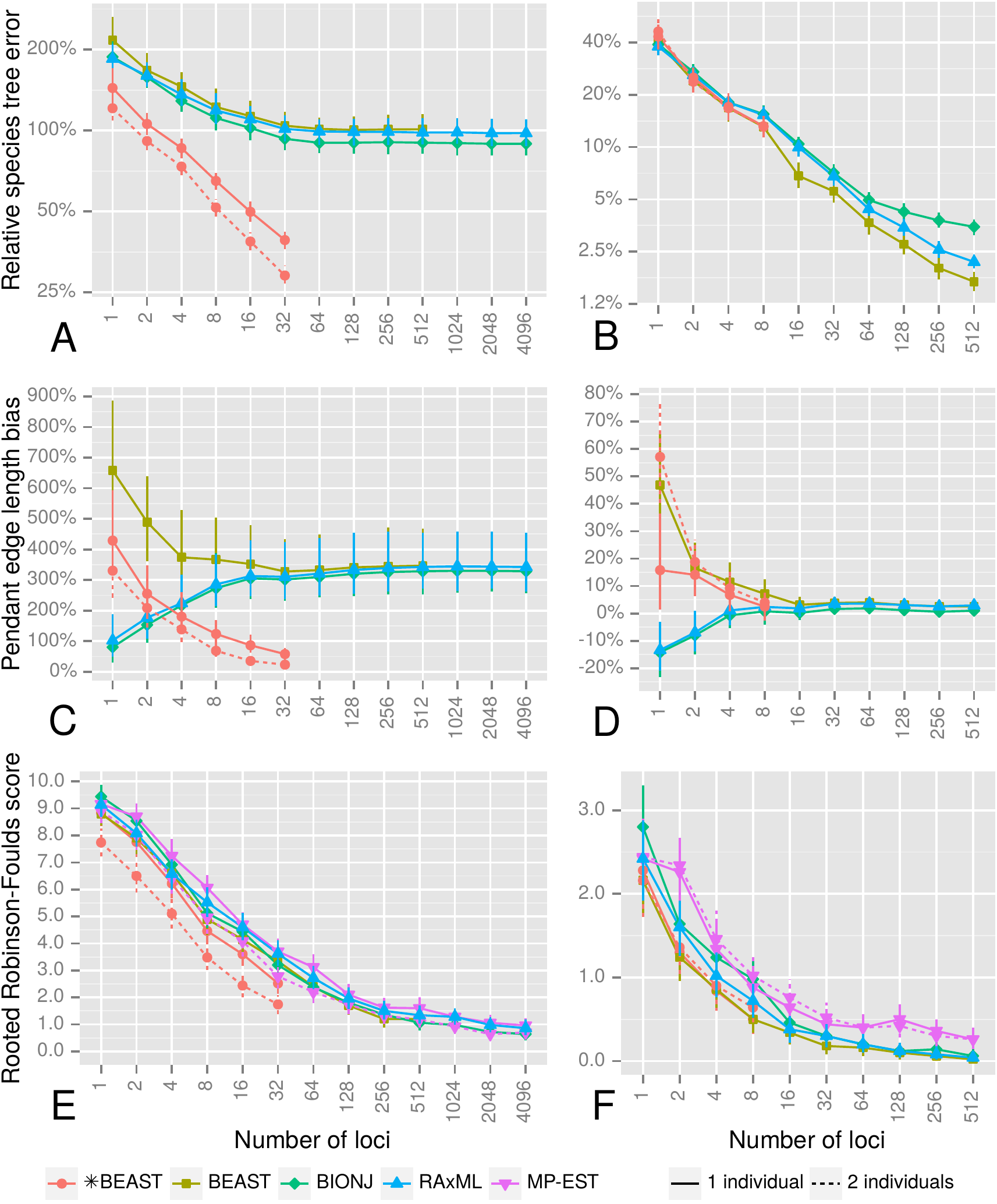}
\caption
{Statistical accuracy of multiple species tree inference methods as a function
of the number of loci. Shallow phylogenetic simulation results (a, c, e) and
deep results (b, d, f) are both presented. Measures of statistical accuracy used
here are relative species tree error (a, b) which incorporates branch length and
topological error, pendant edge length bias (c, d) which highlights biased
branch lengths inferred by non-coalescent methods at the tips of the tree, and
rooted Robinson-Foulds scores (e, f) which are a purely topological measure. All solid
shapes in subfigures a-d show trimmed means (25\% trim to reduce the influence
of outliers), or untrimmed means for subfigures e and f. Vertical range lines show 95\%
confidence intervals for each mean, calculated by bootstrapping.}
\label{fig:multiMethodComparison}
\end{figure}

Based on relative species tree error, *BEAST outperformed all other methods for
any given number of loci for the shallow simulations. The statistical accuracy
of \BayesianSupermatrix{}, RAxML and BIONJ all plateaued beyond 64 loci for the
shallow simulations, whereas *BEAST appears to follow a power law as previously
suggested (Figure~\ref{fig:multiMethodComparison}a). The statistical accuracy of
all methods improves with increasing numbers of loci for the deep simulations,
however we limited the simulations to a maximum of 8 loci when running *BEAST.
BIONJ and RAxML results were similar in accuracy up to 64 loci, but RAxML
outperformed BIONJ by an increasing margin for numbers of loci beyond that
(Figure~\ref{fig:multiMethodComparison}b).

\clearpage

A major factor causing the poor performance of methods other than *BEAST for the
shallow simulations is a bias when estimating pendant edge (also known as leaf
or tip) length. While the mean bias of
estimated pendant edge length trends towards zero in *BEAST, other methods
converge on a bias of approximately 350\%, meaning estimated pendant edges are
on average $4.5\times$ the true length
(Figure~\ref{fig:multiMethodComparison}c). In contrast, there is only a small
positive bias using methods other than *BEAST for the deep simulations
(Figure~\ref{fig:multiMethodComparison}d).

Relative species tree error incorporates both topological error and branch
length error. To separate these two components we calculated the mean rRF score as a
measure of purely topological error --- estimated topologies more distant from the
truth will have higher rRF scores. For shallow simulations, *BEAST was the
best performing method, and the topological accuracy of both *BEAST and \MPEST{} was
improved given two individuals per species (Figure~\ref{fig:multiMethodComparison}e). For
deep simulations, all methods other than *BEAST and \MPEST{} converged at
near-zero topological error given 512 loci (Figure~\ref{fig:multiMethodComparison}f).
*BEAST was limited to a maximum of 8 loci, but its performance for a given number of loci was very close to
\BayesianSupermatrix{}. The topological accuracy of \MPEST{} was inferior to all other methods analyzed.

\section{Discussion and Conclusions}

We have demonstrated by simulation that the multispecies coalescent (as
implemented in *BEAST) can be applied to some problems involving hundreds of
loci. In order to analyze the performance of *BEAST with hundreds of loci under
various conditions, with 100 replicates per condition and given finite
computational resources, we made choices partly based on computational
expediency. These included relatively limited numbers of species and
individuals, and assuming a strict molecular clock. More complexity in the sense
of more parameters to estimate, for example denser taxon sampling or relaxed
clocks, would be expected to require more computational time than the analyses
reported here.

Researchers studying the evolutionary histories of organisms are not burdened by
the need to test hundreds of replicates across many conditions, and can therefore conduct
larger analyses using *BEAST. For example, a recent study of Neotropical
cotingas (Cotingidae: Aves) applied *BEAST to resolve a species tree of 67
extant bird lineages, and used a lognormal relaxed clock for each locus with
molecular rate calibrations to infer absolute divergence times
\citep{Berv2014120}.

\subsection{Power laws describe *BEAST scaling behaviour}

For the various numbers of species, individuals and loci analyzed in this study,
power laws could be used to describe the observed trends in computational
performance of *BEAST, and in the statistical accuracy of the fully Bayesian
multispecies coalescent. In terms of computational performance, this provides a
benchmark for the efficiency of Bayesian MCMC approaches to inference under the
multispecies coalescent. Our results are a product of the particular algorithm
design decisions that the authors of *BEAST have made, and we hope that
power law exponents can be improved upon by subsequent efforts to produce more
efficient algorithms for inference under the multispecies coalescent model.

In contrast, the power law that describes the decrease in estimation uncertainty
associated with inference of the species tree with increasing number of loci is
a fundamental property of the model itself, and will hold regardless of the
details of the algorithmic approach to inference under this model. It therefore
represents a fundamental feature of the problem of species tree inference. With
these results it is possible to extrapolate what one might expect to achieve by
expanding data from a small pilot study to a more comprehensive sample of the
genomic material of a set of study species or individuals.

The decrease in relative species tree error given different numbers of species
and individuals was investigated in experiment 1. Other phylogenetic
parameters were fixed, including the locus length, substitution model and
population size distributions. Possibly because of this, the variation in power
law exponents was minimal. Experiment 3 by contrast compared shallow and deep
phylogenies with larger and smaller population sizes respectively, and
associated alignments of short fixed-length loci and longer variable-length loci
respectively. Clock rate variation and substitution model rates also differed between conditions. Power
law exponents did vary between experiment 1 and both the shallow and deep
inferences in experiment 3; exponents were $-0.433$, $-0.365$ and $-0.568$
respectively. This is important because larger exponents imply a greater decrease
in relative species tree error, so additional loci will lead to a larger
improvement in accuracy of inferred species trees than with a smaller exponent.

Given a hypothetical pilot study of 16 loci, it may be of interest what the
decrease in error would be for a full study of 256 loci. Because the number of
loci in this scenario is increased 16 times, the reduction in relative species
tree error of the full study compared to the pilot study would be $1.0 -
16^{-0.433} \approx 70\%$ if the study is similar to experiment 1, $1.0 -
16^{-0.365} \approx 64\%$ if it is similar to the shallow phylogenetic
simulations, or $1.0 - 16^{-0.568} \approx 79\%$ if it similar to the deep
phylogenetic simulations. What these calculations should
remind us about the power law relationship is that expanding data from 1 to 16
loci provides as great an increase in statistical accuracy as expanding from 16
to 256 loci. That is, for each subsequent locus added there is a diminishing
return with regards to statistical accuracy.

The power laws describing computational performance can also be used to predict
the increase in computational time and chain length required to achieve
sufficient sampling of the posterior distribution. In experiment 1, the power
law coefficient for the log number of loci was $-2.81$ for ESS per hour and
$-1.87$ for ESS per million states. Given the previous example going from 16 to
256 loci, the amount of time required for sufficient sampling of data sets
similar to experiment 1 would increase by $16^{2.81} \approx 2408$ times. The
chain length (number of states) required would increase by $16^{1.87} \approx
180$ times.

Some residual variation in ESS rates was observed after accounting for the
number of individuals, species and loci in each analysis. This was unsurprising
as the operators used by *BEAST are stochastic \citep{Hohna01012012}, so even
when applied to the same data ESS rates are expected to vary between runs.
Consistent with this expectation, the only non-stochastic contribution
identified in our post-hoc analysis was a moderate correlation between residual
ESS per hour and the average gene and species tree height difference.

\subsection{*BEAST compared with other methods}

A previous simulation study which analyzed the scaling behaviour of *BEAST and
other methods used just two species trees to report on topological accuracy given
a range (5, 10, 25 and 50) of number of loci, and produced ambiguous results
\citep{Bayzid:2013vn}. Because we simulated a new species tree for each
replicate, we are able to make more general observations regarding relative
performance. As expected, the relative performance of *BEAST is higher when
branch lengths are shorter. The relative performance of *BEAST is also higher as
the number of loci is increased (Figure~\ref{fig:branchScoreRatio}).

The primary measure we chose to explore statistical accuracy, relative species
tree error, incorporates both branch length and topological error. This measure
is particularly relevant for molecular dating and downstream analyses of
macroevolution and ecology. For example, the PD\textsubscript{C} measure of
phylogenetic diversity and the BiSSE model of binary character influence on
birth and death rates both assume accurate tree topologies and branch lengths
\citep{Maddison01102007,Cadotte04112008}. When inferring species trees with
shorter branch lengths, *BEAST using tens of loci outperformed supermatrix methods by this measure,
even when other methods were able to utilize thousands of loci
(Figure~\ref{fig:multiMethodComparison}a).

If instead branch lengths are irrelevant for a study, *BEAST still outperformed
other methods for a given number of loci when inferring the topology of shallow
species trees (Figure~\ref{fig:multiMethodComparison}e). However when using
thousands of loci other methods were able to outperform *BEAST because
*BEAST was restricted to tens of loci.

For certain species trees concatenation is statistically inconsistent
\citep{Kubatko01022007, Roch201556} and might not outperform *BEAST even when
using thousands of loci. For deeper phylogenetic trees, *BEAST performed
similarly to the \BayesianSupermatrix{} method
(Figure~\ref{fig:multiMethodComparison}b,f).
Because *BEAST requires substantially more computational time,
\BayesianSupermatrix{} methods may be preferable in that case.

Multispecies coalescent methods assume free recombination between loci, and no
recombination within loci. Short sequences dispersed throughout a genome,
including RAD tags, can be justifiably used with coalescent methods as
violations of both assumptions are likely to be limited. However shortcut
coalescence methods like \MPEST{} suffer from high gene tree estimation error when
applied to these short sequences \citep{Mirarab26082014,Springer20161}. In our
study \MPEST{} was inferior to *BEAST and similar to concatenation when inferring
shallow phylogenies using short, RAD tag-like sequences
(Figure~\ref{fig:multiMethodComparison}e). When inferring deep phylogenies
\MPEST{} was inferior to both *BEAST and concatenation
(Figure~\ref{fig:multiMethodComparison}f), despite the longer loci used for
those simulations.

Newer fast multispecies coalescent methods such as ASTRAL
\citep{Mirarab01092014} and SVDquartets \citep{Chifman01122014} may perform
better at inferring species tree topology --- the latest iteration of ASTRAL is
both faster and less sensitive to gene tree error than \MPEST{}
\citep{Mirarab15062015}. However because these methods compute unrooted species
trees without branch lengths, they cannot be compared with other methods using
relative species tree error or rRF scores.

\subsection{Practical implications for applied phylogenetics}

Systematists can use the results of this study as a guide to choosing an
appropriate phylogenetic method. If both \textit{a priori} estimates or
boundaries of root height (clade age) and extant effective population sizes are
available for a particular study system, and the Yule process is a good fit for
that system, an approximate estimate of branch length in coalescent units can be
made before selecting a particular method.

Previous work has shown that the expected mean branch length of a Yule tree is
equal to $\nicefrac{1}{2\lambda}$ \citep{steel2010expected}. Under the Yule model this value is related to the
expected root height:

\begin{equation}
\label{eq:lambda-estimate}
\frac{1}{2\lambda} = \frac{R}{2(H_n - 1)}
\end{equation}

\noindent where $R$ is the expected root height and $H_n$ is the
$n$\textsuperscript{th} harmonic number (where $n$ is the number of species). The
expected branch length $\bar{b}$ in coalescent units of $\tau(2N_e)^{-1}$ is
therefore:

\begin{equation}
\label{eq:branch-length-approximation}
\bar{b} = \frac{1}{2\lambda} \cdot \frac{1}{2N_e} = \frac{1}{4} \cdot \frac{R}{H_n - 1} \cdot \frac{1}{N_e}
\end{equation}

The mean root height of the shallow simulations was 0.01315, and the mean of the
reciprocal extant population sizes $\nicefrac{1}{N_e}$ was 302.05. The
approximate branch length in coalescent units based on these averages is:

\begin{equation}
\label{eq:cyathophora-approximation}
\bar{b} = \frac{1}{4} \cdot \frac{R}{H_n - 1} \cdot \frac{1}{N_e} = \frac{1}{4} \cdot \frac{0.01315}{H_n - 1} \cdot 302.05 = 0.578
\end{equation}

This approximate value is quite close to the sample mean of simulated
branch lengths; $0.539\tau(2N_e)^{-1}$. Based on the results of experiment 2,
this value of $\bar{b}$ is towards the lower bound of the crossover zone, and
*BEAST will be preferred under most conditions
(Figure~\ref{fig:branchScoreRatio}).

The results of experiment 3 will inform researchers with access to phylogenomic
data in the order of hundreds or thousands of loci decide on the appropriate
inference methods. If branch lengths are at all important, either for reporting
divergence times or for downstream analyses which require a species tree, using
a subset of loci with *BEAST will be superior to using all loci with other
methods tested for shallow phylogenies
(Figure~\ref{fig:multiMethodComparison}a). If instead only the topology of the
species tree is of interest, concatenation methods may be superior to fully
Bayesian multispecies coalescent methods like *BEAST until improvements can be
made to their computational performance (Figure~\ref{fig:multiMethodComparison}e,f).

\subsection{Open questions in phylogenomic inference}

Our results point to a number of areas for further research into the
performance of species tree inference.

When using a single locus for species tree inference, experiment 2 shows
\BayesianSupermatrix{} outperforming *BEAST for trees with longer branch
lengths. This may be due to the population size priors used in *BEAST. However
our many-method comparison shows similar performance for both methods given
species trees with long branch lengths. Because deep phylogenetic trees from
experiment 3 were longer than the longest trees from experiment 2, this may
point to a zone of intermediate branch lengths where *BEAST performs poorly
given a single locus.

For all simulations we assumed a constant rate of speciation, however many
lineages of life have undergone rapid radiations. It may be that when inferring
species trees of clades containing ancient rapid radiations the performance of
phylogenetic methods is closer to the shallow simulations than the deep
simulations, and hence *BEAST becomes the preferred method.

Sequence alignments were generated and subsetted uniformly for all simulations
regardless of the number of loci used for each analysis. However researchers may
reasonably choose longer, more informative loci when subsetting phylogenomic
data sets for use with methods like *BEAST which are computationally intensive.
This may improve the relative performance of *BEAST given a subset of the most
informative loci relative to supermatrix or summary methods using thousands of
loci.

However, whole proteins and transcripts can span genomic regions hundreds of thousands of
nucleotides long, so recombination within loci will be common. The use of whole
proteins or transcripts with coalescent methods has been dubbed
``concatalescence'' to reflect this violation \citep{Gatesy26032013,
Gatesy2014231}. If these long sequences are instead split into their constituent
exons, the assumption of free recombination between loci may be violated due to
short intronic distances. Further studies are needed to resolve which violation
is less harmful to statistical accuracy.

\subsection{Conclusion and future directions}

The multispecies coalescent is applicable to a wider range of
conditions then has been suggested by more limited simulation studies. Our
results confirm that the multispecies coalescent is especially suited to
the estimation of shallower evolutionary relationships. We have also
demonstrated that scaling of *BEAST to problems involving
hundreds of loci is feasible, however very long chains and/or crude
parallelization approaches need to be employed.

We anticipate that the increasing availability of phylogenomic sequence data
will motivate further improvements to the computational efficiency of fully
Bayesian inference under the multispecies coalescent model, which should allow
for analysis of hundreds or even thousands of loci across tens or hundreds of
species. These improvements will need to scale efficiently on many-core systems
such as cluster supercomputers, as these systems offer vastly greater computing
power than any desktop workstation.

\section{Funding}

This work was supported by a Rutherford Discovery Fellowship awarded to AJD by
the Royal Society of New Zealand. HAO was supported by an Australian Laureate
Fellowship awarded to Craig Moritz by the Australian Research Council
(FL110100104).

\section{Acknowledgements}

The authors wish to acknowledge the contribution of New Zealand eScience
Infrastructure (NeSI) high-performance computing facilities to the results of
this research, which are funded jointly by NeSI's collaborator institutions and
through the Ministry of Business, Innovation \& Employment's Research
Infrastructure program. This research was undertaken with the assistance of
resources from the National Computational Infrastructure (NCI), which is
supported by the Australian Government. The authors also thank Craig Moritz who
provided valuable suggestions to improve this work.

\clearpage

\bibliographystyle{sysbio}
\bibliography{StarBeastPerformance}

\begin{thebibliography}{68}
\providecommand{\natexlab}[1]{#1}
\providecommand{\selectlanguage}[1]{\relax}
\providecommand{\bibAnnoteFile}[1]{%
  \IfFileExists{#1}{\begin{quotation}\noindent\textsc{Key:} #1\\
  \textsc{Annotation:}\ \input{#1}\end{quotation}}{}}
\providecommand{\bibAnnote}[2]{%
  \begin{quotation}\noindent\textsc{Key:} #1\\
  \textsc{Annotation:}\ #2\end{quotation}}

\bibitem[{Baird et~al.(2008)Baird, Etter, Atwood, Currey, Shiver, Lewis,
  Selker, Cresko, and Johnson}]{Baird:2008ly}
Baird, N.~A., P.~D. Etter, T.~S. Atwood, M.~C. Currey, A.~L. Shiver, Z.~A.
  Lewis, E.~U. Selker, W.~A. Cresko, and E.~A. Johnson. 2008. Rapid {SNP}
  discovery and genetic mapping using sequenced {RAD} markers. PLoS One
  3:e3376.
\bibAnnoteFile{Baird:2008ly}

\bibitem[{Bayzid and Warnow(2013)}]{Bayzid:2013vn}
Bayzid, M.~S. and T.~Warnow. 2013. Naive binning improves phylogenomic
  analyses. Bioinformatics 29:2277--2284.
\bibAnnoteFile{Bayzid:2013vn}

\bibitem[{Benjamini and Hochberg(1995)}]{BenjaminiHochberg1995}
Benjamini, Y. and Y.~Hochberg. 1995. Controlling the false discovery rate: A
  practical and powerful approach to multiple testing. Journal of the Royal
  Statistical Society. Series B (Methodological) 57:289--300.
\bibAnnoteFile{BenjaminiHochberg1995}

\bibitem[{Berv and Prum(2014)}]{Berv2014120}
Berv, J.~S. and R.~O. Prum. 2014. A comprehensive multilocus phylogeny of the
  {Neotropical} cotingas ({Cotingidae}, {Aves}) with a comparative evolutionary
  analysis of breeding system and plumage dimorphism and a revised phylogenetic
  classification. Molecular Phylogenetics and Evolution 81:120--136.
\bibAnnoteFile{Berv2014120}

\bibitem[{Bi et~al.(2012)Bi, Vanderpool, Singhal, Linderoth, Moritz, and
  Good}]{22900609}
Bi, K., D.~Vanderpool, S.~Singhal, T.~Linderoth, C.~Moritz, and J.~Good. 2012.
  Transcriptome-based exon capture enables highly cost-effective comparative
  genomic data collection at moderate evolutionary scales. BMC Genomics 13:403.
\bibAnnoteFile{22900609}

\bibitem[{Bouckaert et~al.(2014)Bouckaert, Heled, K\"uhnert, Vaughan, Wu, Xie,
  Suchard, Rambaut, and Drummond}]{BEAST2}
Bouckaert, R., J.~Heled, D.~K\"uhnert, T.~Vaughan, C.~H. Wu, D.~Xie, M.~A.
  Suchard, A.~Rambaut, and A.~J. Drummond. 2014. {BEAST} 2: A software platform
  for {Bayesian} evolutionary analysis. PLOS Computational Biology 10:e1003537.
\bibAnnoteFile{BEAST2}

\bibitem[{Bryant et~al.(2012)Bryant, Bouckaert, Felsenstein, Rosenberg, and
  RoyChoudhury}]{Bryant01082012}
Bryant, D., R.~Bouckaert, J.~Felsenstein, N.~A. Rosenberg, and A.~RoyChoudhury.
  2012. Inferring species trees directly from biallelic genetic markers:
  Bypassing gene trees in a full coalescent analysis. Molecular Biology and
  Evolution 29:1917--1932.
\bibAnnoteFile{Bryant01082012}

\bibitem[{Cadotte et~al.(2008)Cadotte, Cardinale, and Oakley}]{Cadotte04112008}
Cadotte, M.~W., B.~J. Cardinale, and T.~H. Oakley. 2008. Evolutionary history
  and the effect of biodiversity on plant productivity. Proceedings of the
  National Academy of Sciences 105:17012--17017.
\bibAnnoteFile{Cadotte04112008}

\bibitem[{Camargo et~al.(2012)Camargo, Avila, Morando, and
  Sites}]{camargo2012accuracy}
Camargo, A., L.~J. Avila, M.~Morando, and J.~W. Sites. 2012. Accuracy and
  precision of species trees: effects of locus, individual, and base pair
  sampling on inference of species trees in lizards of the \textit{Liolaemus
  darwinii} group ({Squamata}, {Liolaemidae}). Systematic Biology 61:272--288.
\bibAnnoteFile{camargo2012accuracy}

\bibitem[{Chifman and Kubatko(2014)}]{Chifman01122014}
Chifman, J. and L.~Kubatko. 2014. Quartet inference from {SNP} data under the
  coalescent model. Bioinformatics 30:3317--3324.
\bibAnnoteFile{Chifman01122014}

\bibitem[{Chung and An\'e(2011)}]{chung2011comparing}
Chung, Y. and C.~An\'e. 2011. Comparing two {Bayesian} methods for gene
  tree/species tree reconstruction: Simulations with incomplete lineage sorting
  and horizontal gene transfer. Systematic Biology 60:261--275.
\bibAnnoteFile{chung2011comparing}

\bibitem[{Davey et~al.(2011)Davey, Hohenlohe, Etter, Boone, Catchen, and
  Blaxter}]{Davey:2011zr}
Davey, J.~W., P.~A. Hohenlohe, P.~D. Etter, J.~Q. Boone, J.~M. Catchen, and
  M.~L. Blaxter. 2011. Genome-wide genetic marker discovery and genotyping
  using next-generation sequencing. Nature Reviews Genetics 12:499--510.
\bibAnnoteFile{Davey:2011zr}

\bibitem[{DeGiorgio and Degnan(2010)}]{DeGiorgio:2010ys}
DeGiorgio, M. and J.~H. Degnan. 2010. Fast and consistent estimation of species
  trees using supermatrix rooted triples. Molecular Biology and Evolution
  27:552--569.
\bibAnnoteFile{DeGiorgio:2010ys}

\bibitem[{Eaton and Ree(2013)}]{Eaton01092013}
Eaton, D. A.~R. and R.~H. Ree. 2013. Inferring phylogeny and introgression
  using {RADseq} data: An example from flowering plants {(\textit{Pedicularis}:
  Orobanchaceae)}. Systematic Biology 62:689--706.
\bibAnnoteFile{Eaton01092013}

\bibitem[{Edwards et~al.(2007)Edwards, Liu, and Pearl}]{Edwards03042007}
Edwards, S.~V., L.~Liu, and D.~K. Pearl. 2007. High-resolution species trees
  without concatenation. Proceedings of the National Academy of Sciences
  104:5936--5941.
\bibAnnoteFile{Edwards03042007}

\bibitem[{Faircloth et~al.(2012)Faircloth, McCormack, Crawford, Harvey,
  Brumfield, and Glenn}]{Faircloth:2012ys}
Faircloth, B.~C., J.~E. McCormack, N.~G. Crawford, M.~G. Harvey, R.~T.
  Brumfield, and T.~C. Glenn. 2012. Ultraconserved elements anchor thousands of
  genetic markers spanning multiple evolutionary timescales. Systematic Biology
  61:717--726.
\bibAnnoteFile{Faircloth:2012ys}

\bibitem[{Gascuel(1997)}]{Gascuel01071997}
Gascuel, O. 1997. {BIONJ}: An improved version of the {NJ} algorithm based on a
  simple model of sequence data. Molecular Biology and Evolution 14:685--695.
\bibAnnoteFile{Gascuel01071997}

\bibitem[{Gatesy and Springer(2013)}]{Gatesy26032013}
Gatesy, J. and M.~S. Springer. 2013. Concatenation versus coalescence versus
  ``concatalescence''. Proceedings of the National Academy of Sciences
  110:E1179.
\bibAnnoteFile{Gatesy26032013}

\bibitem[{Gatesy and Springer(2014)}]{Gatesy2014231}
Gatesy, J. and M.~S. Springer. 2014. Phylogenetic analysis at deep timescales:
  Unreliable gene trees, bypassed hidden support, and the
  coalescence/concatalescence conundrum. Molecular Phylogenetics and Evolution
  80:231--266.
\bibAnnoteFile{Gatesy2014231}

\bibitem[{Gernhard(2008)}]{Gernhard:2008uq}
Gernhard, T. 2008. The conditioned reconstructed process. Journal of
  Theoretical Biology 253:769--778.
\bibAnnoteFile{Gernhard:2008uq}

\bibitem[{Hasegawa et~al.(1985)Hasegawa, Kishino, and Yano}]{Hasegawa1985}
Hasegawa, M., H.~Kishino, and T.~Yano. 1985. Dating the human-ape splitting by
  a molecular clock of mitochondrial dna. Journal of Molecular Evolution
  22:160--174.
\bibAnnoteFile{Hasegawa1985}

\bibitem[{Heled and Bouckaert(2013)}]{24093883}
Heled, J. and R.~Bouckaert. 2013. Looking for trees in the forest: Summary tree
  from posterior samples. BMC Evolutionary Biology 13:221.
\bibAnnoteFile{24093883}

\bibitem[{Heled et~al.(2013)Heled, Bryant, and Drummond}]{Heled:2013fk}
Heled, J., D.~Bryant, and A.~J. Drummond. 2013. Simulating gene trees under the
  multispecies coalescent and time-dependent migration. BMC Evolutionary
  Biology 13:44.
\bibAnnoteFile{Heled:2013fk}

\bibitem[{Heled and Drummond(2010)}]{*BEAST}
Heled, J. and A.~Drummond. 2010. Bayesian inference of species trees from
  multilocus data. Molecular Biology and Evolution 27:570--580.
\bibAnnoteFile{*BEAST}

\bibitem[{H\"ohna and Drummond(2012)}]{Hohna01012012}
H\"ohna, S. and A.~J. Drummond. 2012. Guided tree topology proposals for
  {Bayesian} phylogenetic inference. Systematic Biology 61:1--11.
\bibAnnoteFile{Hohna01012012}

\bibitem[{Joly et~al.(2009)Joly, McLenachan, and Lockhart}]{Joly2009}
Joly, S., P.~A. McLenachan, and P.~J. Lockhart. 2009. A statistical approach
  for distinguishing hybridization and incomplete lineage sorting. The American
  Naturalist 174:E54--E70.
\bibAnnoteFile{Joly2009}

\bibitem[{Jukes and Cantor(1969)}]{Jukes1969}
Jukes, T. and C.~Cantor. 1969. Evolution of protein molecules. Pages~21--132
  \emph{in} Mammalian Protein Metabolism (H.~Munro, ed.). Academic Press, New
  York.
\bibAnnoteFile{Jukes1969}

\bibitem[{Kendall(1948)}]{Kendall1948b}
Kendall, D.~G. 1948. On the generalized ``birth-and-death'' process. Annals of
  Mathematical Statistics 19:1--15.
\bibAnnoteFile{Kendall1948b}

\bibitem[{Kirkpatrick and Slatkin(1993)}]{KirkpatrickSlatkin1993}
Kirkpatrick, M. and M.~Slatkin. 1993. Searching for evolutionary patterns in
  the shape of a phylogenetic tree. Evolution 47:1171--1181.
\bibAnnoteFile{KirkpatrickSlatkin1993}

\bibitem[{Kubatko(2009)}]{Kubatko01102009}
Kubatko, L.~S. 2009. Identifying hybridization events in the presence of
  coalescence via model selection. Systematic Biology 58:478--488.
\bibAnnoteFile{Kubatko01102009}

\bibitem[{Kubatko et~al.(2009)Kubatko, Carstens, and Knowles}]{Kubatko:2009qf}
Kubatko, L.~S., B.~C. Carstens, and L.~L. Knowles. 2009. {STEM}: Species tree
  estimation using maximum likelihood for gene trees under coalescence.
  Bioinformatics 25:971--973.
\bibAnnoteFile{Kubatko:2009qf}

\bibitem[{Kubatko and Degnan(2007)}]{Kubatko01022007}
Kubatko, L.~S. and J.~H. Degnan. 2007. Inconsistency of phylogenetic estimates
  from concatenated data under coalescence. Systematic Biology 56:17--24.
\bibAnnoteFile{Kubatko01022007}

\bibitem[{Lanier et~al.(2014)Lanier, Huang, and Knowles}]{Lanier:2014kx}
Lanier, H.~C., H.~Huang, and L.~L. Knowles. 2014. How low can you go? {The}
  effects of mutation rate on the accuracy of species-tree estimation.
  Molecular Phylogenetics and Evolution 70:112--119.
\bibAnnoteFile{Lanier:2014kx}

\bibitem[{Larget et~al.(2010)Larget, Kotha, Dewey, and An{\'e}}]{Larget:2010zr}
Larget, B.~R., S.~K. Kotha, C.~N. Dewey, and C.~An{\'e}. 2010. {BUCKy}: Gene
  tree/species tree reconciliation with {Bayesian} concordance analysis.
  Bioinformatics 26:2910--2911.
\bibAnnoteFile{Larget:2010zr}

\bibitem[{Leach\'e et~al.(2014)Leach\'e, Harris, Rannala, and
  Yang}]{Leache01012014}
Leach\'e, A.~D., R.~B. Harris, B.~Rannala, and Z.~Yang. 2014. The influence of
  gene flow on species tree estimation: A simulation study. Systematic Biology
  63:17--30.
\bibAnnoteFile{Leache01012014}

\bibitem[{Leach{\'e} and Rannala(2011)}]{leache2011accuracy}
Leach{\'e}, A.~D. and B.~Rannala. 2011. The accuracy of species tree estimation
  under simulation: a comparison of methods. Systematic Biology 60:126--137.
\bibAnnoteFile{leache2011accuracy}

\bibitem[{Lemmon et~al.(2012)Lemmon, Emme, and Lemmon}]{Lemmon:2012fk}
Lemmon, A.~R., S.~A. Emme, and E.~M. Lemmon. 2012. Anchored hybrid enrichment
  for massively high-throughput phylogenomics. Systematic Biology 61:727--744.
\bibAnnoteFile{Lemmon:2012fk}

\bibitem[{Liu(2008)}]{liu2008best}
Liu, L. 2008. {BEST}: {Bayesian} estimation of species trees under the
  coalescent model. Bioinformatics 24:2542--2543.
\bibAnnoteFile{liu2008best}

\bibitem[{Liu et~al.(2008)Liu, Pearl, Brumfield, and Edwards}]{Liu:2008kx}
Liu, L., D.~K. Pearl, R.~T. Brumfield, and S.~V. Edwards. 2008. Estimating
  species trees using multiple-allele {DNA} sequence data. Evolution
  62:2080--2091.
\bibAnnoteFile{Liu:2008kx}

\bibitem[{Liu et~al.(2010)Liu, Yu, and Edwards}]{20937096}
Liu, L., L.~Yu, and S.~Edwards. 2010. A maximum pseudo-likelihood approach for
  estimating species trees under the coalescent model. BMC Evolutionary Biology
  10:302.
\bibAnnoteFile{20937096}

\bibitem[{Liu et~al.(2009)Liu, Yu, Pearl, and Edwards}]{Liu:2009ve}
Liu, L., L.~Yu, D.~K. Pearl, and S.~V. Edwards. 2009. Estimating species
  phylogenies using coalescence times among sequences. Systematic Biology
  58:468--477.
\bibAnnoteFile{Liu:2009ve}

\bibitem[{Maddison(1997)}]{Maddison01091997}
Maddison, W.~P. 1997. Gene trees in species trees. Systematic Biology
  46:523--536.
\bibAnnoteFile{Maddison01091997}

\bibitem[{Maddison et~al.(2007)Maddison, Midford, and Otto}]{Maddison01102007}
Maddison, W.~P., P.~E. Midford, and S.~P. Otto. 2007. Estimating a binary
  character's effect on speciation and extinction. Systematic Biology
  56:701--710.
\bibAnnoteFile{Maddison01102007}

\bibitem[{Mamanova et~al.(2010)Mamanova, Coffey, Scott, Kozarewa, Turner,
  Kumar, Howard, Shendure, and Turner}]{Mamanova:2010vn}
Mamanova, L., A.~J. Coffey, C.~E. Scott, I.~Kozarewa, E.~H. Turner, A.~Kumar,
  E.~Howard, J.~Shendure, and D.~J. Turner. 2010. Target-enrichment strategies
  for next-generation sequencing. Nature Methods 7:111--118.
\bibAnnoteFile{Mamanova:2010vn}

\bibitem[{McCormack et~al.(2013)McCormack, Hird, Zellmer, Carstens, and
  Brumfield}]{McCormack:2013kx}
McCormack, J.~E., S.~M. Hird, A.~J. Zellmer, B.~C. Carstens, and R.~T.
  Brumfield. 2013. Applications of next-generation sequencing to phylogeography
  and phylogenetics. Molecular Phylogenetics and Evolution 66:526--538.
\bibAnnoteFile{McCormack:2013kx}

\bibitem[{Mirarab et~al.(2014{\natexlab{a}})Mirarab, Bayzid, and
  Warnow}]{Mirarab26082014}
Mirarab, S., M.~S. Bayzid, and T.~Warnow. 2014{\natexlab{a}}. Evaluating
  summary methods for multilocus species tree estimation in the presence of
  incomplete lineage sorting. Systematic Biology published online: August 26,
  2014.
\bibAnnoteFile{Mirarab26082014}

\bibitem[{Mirarab et~al.(2014{\natexlab{b}})Mirarab, Reaz, Bayzid, Zimmermann,
  Swenson, and Warnow}]{Mirarab01092014}
Mirarab, S., R.~Reaz, M.~S. Bayzid, T.~Zimmermann, M.~S. Swenson, and
  T.~Warnow. 2014{\natexlab{b}}. {ASTRAL}: Genome-scale coalescent-based
  species tree estimation. Bioinformatics 30:i541--i548.
\bibAnnoteFile{Mirarab01092014}

\bibitem[{Mirarab and Warnow(2015)}]{Mirarab15062015}
Mirarab, S. and T.~Warnow. 2015. {ASTRAL-II}: Coalescent-based species tree
  estimation with many hundreds of taxa and thousands of genes. Bioinformatics
  31:i44--i52.
\bibAnnoteFile{Mirarab15062015}

\bibitem[{Nee et~al.(1994)Nee, Holmes, May, and Harvey}]{Nee:1994fk}
Nee, S., E.~C. Holmes, R.~M. May, and P.~H. Harvey. 1994. Extinction rates can
  be estimated from molecular phylogenies. Philosophical Transactions of the
  Royal Society B: Biological Sciences 344:77--82.
\bibAnnoteFile{Nee:1994fk}

\bibitem[{Page and Charleston(1997)}]{Page1997231}
Page, R. D.~M. and M.~A. Charleston. 1997. From gene to organismal phylogeny:
  Reconciled trees and the gene tree/species tree problem. Molecular
  Phylogenetics and Evolution 7:231--240.
\bibAnnoteFile{Page1997231}

\bibitem[{Pamilo and Nei(1988)}]{Pamilo:1988fk}
Pamilo, P. and M.~Nei. 1988. Relationships between gene trees and species
  trees. Molecular Biology and Evolution 5:568--583.
\bibAnnoteFile{Pamilo:1988fk}

\bibitem[{Perry et~al.(2012)Perry, Melsted, Marioni, Wang, Bainer, Pickrell,
  Michelini, Zehr, Yoder, Stephens, Pritchard, and Gilad}]{Perry01042012}
Perry, G.~H., P.~Melsted, J.~C. Marioni, Y.~Wang, R.~Bainer, J.~K. Pickrell,
  K.~Michelini, S.~Zehr, A.~D. Yoder, M.~Stephens, J.~K. Pritchard, and
  Y.~Gilad. 2012. Comparative {RNA} sequencing reveals substantial genetic
  variation in endangered primates. Genome Research 22:602--610.
\bibAnnoteFile{Perry01042012}

\bibitem[{Rambaut and Grass(1997)}]{Rambaut01061997}
Rambaut, A. and N.~C. Grass. 1997. {Seq-Gen}: an application for the {Monte
  Carlo} simulation of {DNA} sequence evolution along phylogenetic trees.
  Computer Applications in the Biosciences 13:235--238.
\bibAnnoteFile{Rambaut01061997}

\bibitem[{Rannala and Yang(2003)}]{RannalaYang2003}
Rannala, B. and Z.~Yang. 2003. Bayes estimation of species divergence times and
  ancestral population sizes using {DNA} sequences from multiple loci. Genetics
  164:1645--1656.
\bibAnnoteFile{RannalaYang2003}

\bibitem[{Roch and Steel(2015)}]{Roch201556}
Roch, S. and M.~Steel. 2015. Likelihood-based tree reconstruction on a
  concatenation of aligned sequence data sets can be statistically
  inconsistent. Theoretical Population Biology 100:56--62.
\bibAnnoteFile{Roch201556}

\bibitem[{Slowinski and Page(1999)}]{Slowinski01101999}
Slowinski, J.~B. and R.~D.~M. Page. 1999. How should species phylogenies be
  inferred from sequence data? Systematic Biology 48:814--825.
\bibAnnoteFile{Slowinski01101999}

\bibitem[{Springer and Gatesy(2016)}]{Springer20161}
Springer, M.~S. and J.~Gatesy. 2016. The gene tree delusion. Molecular
  Phylogenetics and Evolution 94, Part A:1 -- 33.
\bibAnnoteFile{Springer20161}

\bibitem[{Stamatakis(2014)}]{Stamatakis01052014}
Stamatakis, A. 2014. {RAxML} version 8: A tool for phylogenetic analysis and
  post-analysis of large phylogenies. Bioinformatics 30:1312--1313.
\bibAnnoteFile{Stamatakis01052014}

\bibitem[{Steel and Mooers(2010)}]{steel2010expected}
Steel, M. and A.~Mooers. 2010. The expected length of pendant and interior
  edges of a yule tree. Applied Mathematics Letters 23:1315--1319.
\bibAnnoteFile{steel2010expected}

\bibitem[{Steiper and Young(2006)}]{Steiper2006384}
Steiper, M.~E. and N.~M. Young. 2006. Primate molecular divergence dates.
  Molecular Phylogenetics and Evolution 41:384--394.
\bibAnnoteFile{Steiper2006384}

\bibitem[{Sukumaran and Holder(2010)}]{Sukumaran15062010}
Sukumaran, J. and M.~T. Holder. 2010. {DendroPy}: a {Python} library for
  phylogenetic computing. Bioinformatics 26:1569--1571.
\bibAnnoteFile{Sukumaran15062010}

\bibitem[{Tavar\'e et~al.(2002)Tavar\'e, Marshall, Will, Soligo, and
  Martin}]{Tavare2002}
Tavar\'e, S., C.~R. Marshall, O.~Will, C.~Soligo, and R.~D. Martin. 2002. Using
  the fossil record to estimate the age of the last common ancestor of extant
  primates. Nature 416:726--729.
\bibAnnoteFile{Tavare2002}

\bibitem[{Wiens and Morrill(2011)}]{Wiens:2011uq}
Wiens, J.~J. and M.~C. Morrill. 2011. Missing data in phylogenetic analysis:
  Reconciling results from simulations and empirical data. Systematic Biology
  60:719--731.
\bibAnnoteFile{Wiens:2011uq}

\bibitem[{Wilkinson et~al.(2011)Wilkinson, Steiper, Soligo, Martin, Yang, and
  Tavar\'e}]{Wilkinson2011}
Wilkinson, R.~D., M.~E. Steiper, C.~Soligo, R.~D. Martin, Z.~Yang, and
  S.~Tavar\'e. 2011. Dating primate divergences through an integrated analysis
  of palaeontological and molecular data. Systematic Biology 60:16--31.
\bibAnnoteFile{Wilkinson2011}

\bibitem[{Yang and Wang(2007)}]{YangWang2007}
Yang, F.~S. and X.~Q. Wang. 2007. Extensive length variation in the {cpDNA
  \textit{trn}T-\textit{trn}F} region of hemiparasitic \textit{Pedicularis} and
  its phylogenetic implications. Plant Systematics and Evolution 264:251--264.
\bibAnnoteFile{YangWang2007}

\bibitem[{Yang et~al.(1994)Yang, Goldman, and Friday}]{Yang01031994}
Yang, Z., N.~Goldman, and A.~Friday. 1994. Comparison of models for nucleotide
  substitution used in maximum-likelihood phylogenetic estimation. Molecular
  Biology and Evolution 11:316--324.
\bibAnnoteFile{Yang01031994}

\bibitem[{Yu et~al.(2011)Yu, Than, Degnan, and Nakhleh}]{yu2011coalescent}
Yu, Y., C.~Than, J.~H. Degnan, and L.~Nakhleh. 2011. Coalescent histories on
  phylogenetic networks and detection of hybridization despite incomplete
  lineage sorting. Systematic Biology 60:138--149.
\bibAnnoteFile{yu2011coalescent}

\bibitem[{Yule(1924)}]{Yule1924}
Yule, G.~U. 1924. A mathematical theory of evolution based on the conclusions
  of {Dr.~J.C.~Willis}. Philosophical Transactions of the Royal Society B:
  Biological Sciences 213:21--87.
\bibAnnoteFile{Yule1924}

\end{thebibliography}

\end{document}